\documentclass[12pt]{article}

\usepackage{amsmath}
\usepackage{amsfonts}
\usepackage{amssymb}
\usepackage{graphics,graphicx,tikz}
\usepackage{subfig}
\usepackage{graphicx}
\usepackage{soul}
\usepackage{cite}
\usepackage{graphicx}
\usetikzlibrary{patterns}
\usepackage{enumitem}
\usepackage{array}
\usepackage{amsbsy}
\usepackage{bbold}
\usepackage{appendix}
\usepackage{mathrsfs}
\usepackage{physics}
\usepackage{amsthm}
\usepackage{cancel}
\usepackage{yfonts}
\usepackage{inputenc}
\usepackage{chngcntr}

\numberwithin{equation}{section} 
\usepackage[linktocpage]{hyperref}
\hypersetup{
	colorlinks=true,
	linkcolor=blue,
	filecolor=magenta,      
	urlcolor=blue,
	citecolor=blue
}
\def\beq{\begin{equation}}
\def\eeq{\end{equation}}
\newcommand{\commentOut}[1]{}
\def\bea{\begin{align}}
\def\eea{\end{align}}

\usepackage{color}

\begin{document}

\begin{titlepage}
\begin{flushright}
\end{flushright}
\vskip 1.0cm
\begin{center}
{\Large \bf 
Logarithmic soft theorems and soft spectra
}

\vskip 1.0cm {\large  Francesco Alessio$^{a,b}$, Paolo Di Vecchia$^{a, c}$, Carlo Heissenberg$^d$} \\[0.7cm]

{\it \small $^a$NORDITA, KTH Royal Institute of Technology and Stockholm University, \\ Hannes Alfv{\'{e}}ns v{\"{a}}g 12, SE-11419 Stockholm, Sweden  }\\
{\it \small $^b$Department of Physics and Astronomy, Uppsala University,\\ Box 516, SE-75120 Uppsala, Sweden}

{\it \small $^c$The Niels Bohr Institute, Blegdamsvej 17, DK-2100 Copenhagen, Denmark}\\

{\it \small $^d$Queen Mary University of London, School of Mathematical Sciences,\\
Mile End Road, E1 4NS, United Kingdom}\\

\end{center}

\begin{abstract}
Using universal predictions provided by classical soft theorems, we revisit the energy emission spectrum for gravitational scatterings of compact objects in the low-frequency expansion. We calculate this observable beyond the zero-frequency limit, retaining an exact dependence on the kinematics of the massive objects. This allows us to study independently the ultrarelativistic or massless limit, where we find agreement with the literature, and the small-deflection or post-Minkowskian (PM) limit, where we provide explicit results up to $\mathcal{O}(G^5)$. These  confirm that the high-velocity limit of a given PM order is smoothly connected to the corresponding massless result whenever the latter is analytic in the Newton constant $G$. We also provide explicit expressions for the
waveforms to order $\omega^{-1}$, $\log\omega$, $\omega(\log\omega)^2$ in the soft limit, $\omega\to0$,
expanded up to sub-subleading PM order,
as well as a conjecture for the logarithmic soft terms of the type $\omega^{n-1}(\log\omega)^{n}$ with $n\ge 3$.

\end{abstract}

\end{titlepage}

\tableofcontents

\section{Introduction}

Classical soft theorems provide universal constraints that dictate the nonanalytic behavior of electromagnetic and gravitational waveforms in the low-frequency regime, or equivalently, via Fourier transform, at early and late times. 
Perhaps the most well-known prediction of this kind is the gravitational memory effect \cite{Zeldovich:1974gvh,Christodoulou:1991cr,Wiseman:1991ss,Thorne:1992sdb}, which determines the permanent change in the asymptotic metric fluctuation induced by the passage of a gravitational wave, or equivalently the leading $\frac{1}{\omega}$ pole in frequency domain \cite{Weinberg:1964ew,Weinberg:1965nx,Strominger:2014pwa}. The physical intuition at the basis of this result is that the leading contributions due to very soft gravitons arise due to emissions from the initial and final massive states (\emph{linear} memory) or from the emitted gravitons themselves (\emph{nonlinear} memory). The connections between soft graviton theorems, memory effects and asymptotic symmetries have been identified as an infrared triangle that characterizes key properties of gauge theories and gravity in the infrared \cite{Strominger:2013jfa,He:2014laa,He:2014cra,Strominger:2014pwa,Campiglia:2015qka,Kapec:2015ena,Strominger:2017zoo}.

The power of soft theorems in identifying constraints for low-frequency waveforms was emphasized in particular in a series of papers \cite{Sahoo:2018lxl,Laddha:2019yaj,Saha:2019tub,Sahoo:2020ryf,Sahoo:2021ctw,Ghosh:2021bam}, where the authors obtained general predictions for the waveforms beyond the leading $\frac{1}{\omega}$ pole that also dictate the ``leading logarithms'' $\log\omega$, corresponding to a $\frac{1}{|t|}$ tail in time domain, and $\omega(\log\omega)^2$, corresponding to $\frac{1}{t^2}\,\log|t|$. These general expressions are simple functions of the initial and final momenta of the objects that take part in the collision and thus source the signal. 
In particular, it was noted in \cite{Sahoo:2021ctw} that, while the nonlinear memory effect induces nontrivial contributions due to soft emissions from final gravitons in the $\frac{1}{\omega}$ term, subleading $\log\omega$ and subsubleading $\omega(\log\omega)^2$ terms afford elementary expressions that only involve the massive momenta, owing to nontrivial cancellations. 
As described in \cite{Krishna:2023fxg,Agrawal:2023zea,Alessio:2024wmz} (see also \cite{Choi:2024ygx}), such cancellations can be traced back to the universality of the first three soft factors, whose expressions are completely fixed in a simple way by the structure of infrared divergences.

In this work, we use these universal soft factors expressed in terms of initial and final massive momenta to calculate the soft energy emission spectrum, $\frac{dE}{d\omega}$ as $\omega\to0$, which characterizes the amount of energy that a gravitational system loses via the emission of low-frequency gravitational waves.
To leading order, this function reduces to an $\omega^0$ value often referred to as Zero Frequency Limit (ZFL), whose dependence on the initial and final momenta is fixed by the leading $\frac{1}{\omega}$ soft theorem \cite{Weinberg:1965nx,DiVecchia:2022nna}.
The main new result of this paper is to obtain an analogous expression also for the Next-to-Leading Order (NLO) contribution, which scales as $\omega^2(\log\omega)^2$ for small $\omega$. In contrast with the ZFL, this new term also involves the $\log\omega$ and $\omega(\log\omega)^2$ soft factors for the corresponding waveform. A novel feature of this result is that, while the ZFL can be expressed as a function of   Lorentz-invariant products of initial and final momenta, the NLO inherits from the definition of the frequency an explicit dependence on the observer's frame, so that for instance its expression will differ when considering the center-of-mass frame or the one where a given particle is initially at rest. 

Soft theorems, and the resulting expressions for the ZFL and the NLO contribution to the spectrum, retain an exact dependence on the kinematics of the hard objects. It is then very instructive to explicitly study their behavior in various kinematic limits. 
In the ultrarelativistic regime, that is, when the colliding objects move very fast and their masses become negligible, our results reduce to the ones already obtained for scatterings of massless objects in \cite{Gruzinov:2014moa,Ciafaloni:2015vsa,Ciafaloni:2015xsr,Ciafaloni:2018uwe,Sahoo:2021ctw}. The small-deflection regime corresponds instead to the situation in which the colliding objects are well separated, and gravitational interactions are weak, so that one makes contact with the post-Minkowskian (PM) expansion. 

This approximation, whose natural power-counting parameter is the Newton constant $G$, has received recently a great deal of attention due to the realization that PM observables can be extracted in a simple way from the perturbative expansion of scattering amplitudes \cite{Bjerrum-Bohr:2018xdl,Kosower:2018adc,Bern:2019crd,Bern:2019nnu,Bern:2021dqo,Bern:2021yeh,Damgaard:2023ttc}. 
Progress in this direction has encompassed both the calculation of the deflection that two massive objects experience in a classical gravitational scattering \cite{Bern:2019nnu,KoemansCollado:2019ggb,Bern:2019crd,Cristofoli:2020uzm,Parra-Martinez:2020dzs,DiVecchia:2020ymx,DiVecchia:2021ndb,Herrmann:2021lqe,DiVecchia:2021bdo,Herrmann:2021tct,Bjerrum-Bohr:2021vuf,Heissenberg:2021tzo,Bjerrum-Bohr:2021din,Damgaard:2021ipf,Brandhuber:2021eyq,Bern:2021dqo,Bern:2021yeh,Dlapa:2022lmu,Dlapa:2023hsl,Damgaard:2023ttc,Driesse:2024xad,Bern:2024adl} and the gravitational waveform itself \cite{Goldberger:2016iau,Luna:2017dtq,Mogull:2020sak,Jakobsen:2021smu,Mougiakakos:2021ckm,Brandhuber:2023hhy,Herderschee:2023fxh,Elkhidir:2023dco,Georgoudis:2023lgf,Caron-Huot:2023vxl,DeAngelis:2023lvf,Brandhuber:2023hhl,Aoude:2023dui,Georgoudis:2023eke,Georgoudis:2023ozp,Bohnenblust:2023qmy,Adamo:2024oxy,Georgoudis:2024pdz,Bini:2024rsy,Bini:2024ijq}. See also \cite{Kalin:2020mvi,Kalin:2020fhe,Mogull:2020sak,Dlapa:2021npj,Jakobsen:2022psy,Kalin:2022hph,Dlapa:2022lmu,Dlapa:2023hsl,Driesse:2024xad} and references therein for powerful techniques to solve the two-body problem by worldline methods.

Focusing on the PM regime of a $2\to2$ gravitational process, i.e.~the expansion for small momentum transfer, we provide explicit expressions for the ZFL and the NLO of the spectrum $\frac{dE}{d\omega}$ that are valid up to $\mathcal{O}(G^5)$, both in the center-of-mass frame and in the rest frame of either particle. The first nontrivial contribution, $\mathcal{O}(G^3)$, corresponds to the soft limit of a cut two-loop calculation, whose integral over all frequencies would lead to the total emitted energy obtained in \cite{Herrmann:2021lqe,Herrmann:2021tct}.
The $\mathcal{O}(G^4)$ term of the NLO, $\omega^2(\log\omega)^2$, piece turns out to vanish, while the $\mathcal{O}(G^5)$ would correspond to a four-loop calculation on the amplitude side.
In a similar spirit, we also provide the explicit $G$-expansion of the soft theorems for the waveform, which extend the subleading PM results cross-checked in \cite{Bini:2023fiz,Georgoudis:2023eke} against explicit calculations, and can serve as consistency checks for subleading PM calculations at two loops (see also \cite{Bautista:2021llr}).
While these results are valid for generic velocities, we also consider their limit in which the objects' velocities are small, where we can compare with available post-Newtonian (PN) results \cite{Bini:2021jmj,Bini:2023fiz,Georgoudis:2024pdz,Bini:2024rsy} finding perfect agreement with our formulas.
Based on patterns that emerge in the expressions, we also propose a conjecture for the general form of the soft factor dictating the coefficient of the leading logarithmic terms $\omega^{n-1}(\log\omega)^n$ for $n\ge3$ in the small-frequency expansion of the waveform. We find that this proposal becomes particularly simple for the case of a $2\to2$ hard process and in this case it passes a simple but nontrivial check in the Newtonian (0PN) limit, where we provide an explicit derivation of such leading logarithms from the Keplerian trajectory. It also agrees with the recent results in \cite{Bini:2024ijq} at sub-subleading order in the PM expansion and up to 2PN order. The proposed all-order formulas allow us to derive resummed expressions for the waveform that capture all leading logarithms, and which we later use to provide similar resummed predictions for the spectrum for the $2\to2$ case. Our conjecture was also cross-checked with results obtained to leading order in the near-probe limit in Ref.~\cite{Fucito:2024wlg} for an $N\to N$ hard process in which one object is much heavier than the remaining $N-1$. See Eqs.~\eqref{eq:finalproposal2to2} and Eqs.~\eqref{eq:finalproposalNtoM} below for the conjectured resummed waveforms in the $2\to2$ and $N\to M$ cases.

Both for the ZFL and for the NLO of the energy spectrum, the validity of our results is limited to $\mathcal{O}(G^5)$ because at $\mathcal{O}(G^6)$ the nonlinear memory part of the waveform will start contributing. In fact, interference terms involving nonlinear memory can in principle show up in the ZFL already at $\mathcal{O}(G^5)$ by power counting; however, we shall see that a parity property of the leading-order spectral density of graviton emissions ensures that such a contribution vanishes identically, as already noted in \cite{DiVecchia:2022nna} in the PN limit (see \cite{Heissenberg:2024umh} for a similar mechanism at play in the $\mathcal{O}(G^4)$ angular momentum loss). More generally, for $n\ge1$, possible nonlinear memory corrections limit our predictions for the $(\omega\log\omega)^{2n}$ piece of the spectrum to be valid up to and including $\mathcal{O}(G^{2n+3})$.

As in Ref.~\cite{DiVecchia:2022nna}, one of our main motivations for obtaining explicit expressions of low-frequency observables is to be able to discuss the relation between the ``true'' ultrarelativistic limit, performed for generic deflections, and the large-velocity limit of each PM contribution. It was realized that, after the addition of radiation reaction, the 3PM deflection angle \cite{Bern:2019crd,Bern:2019nnu,DiVecchia:2020ymx,Damour:2020tta,DiVecchia:2021ndb,DiVecchia:2021bdo,Alessio:2022kwv} admits a smooth high-energy limit, in which it reduces to the classic result of Ref.~\cite{Amati:1990xe} for the massless case. However, this is not the case for the ZFL \cite{DiVecchia:2022nna}, where the high-energy limit of the $\mathcal{O}(G^3)$ result seemingly diverges. This unphysical feature can be ascribed to the fact that the correct ultrarelativistic limit \cite{Gruzinov:2014moa,Ciafaloni:2015xsr,Ciafaloni:2015vsa,Sahoo:2021ctw} is nonanalytic, and the resulting branch cut dictates a finite radius of convergence of the PM series. A similar mechanism also features in the zero-frequency contribution to the angular momentum loss \cite{DiVecchia:2022owy,Heissenberg:2024umh}.
In the NLO contribution to the energy spectrum that we obtain in this paper, a similar pattern emerges. The $\mathcal{O}(G^3)$ admits a smooth high-energy limit, which matches the ultrarelativistic result re-expanded for small deflections, while the $\mathcal{O}(G^5)$ exhibits a logarithmic singularity, corresponding to the appearance of a nonanalytic logarithmic dependence on $G$ in the massless result. We find that this pattern continues for all leading and subleading PM contribution to all leading logs (see Eq.~\eqref{eq:resummedLeadingLogs} below). It would be interesting to elucidate the analogous mechanism at the basis of similar (albeit power-like) high-velocity singularities in the total radiated energy-momentum at $\mathcal{O}(G^3)$ \cite{Herrmann:2021lqe,Herrmann:2021tct} and of the deflection angle at $\mathcal{O}(G^4)$ \cite{Dlapa:2022lmu,Damgaard:2023ttc}.

Based on the explicit expressions for the $\omega^0$ and $\omega \log\omega$ tree-level waveforms obtained in \cite{Georgoudis:2023eke} for the scattering of minimally coupled massive scalars, we are also able to calculate the corresponding $\omega^2 \log\omega$ contribution to the $\mathcal{O}(G^3)$ energy spectrum. Once again, we perform the calculation in two different frames, which highlights in particular a very different dependence on the symmetric mass-ratio depending on the observer under consideration. 

The structure of the paper is as follows. In Section~\ref{sec:lowfreq}, we review the classical soft graviton theorems, highlighting the role of nonlinear memory contributions and how they can be ``rewritten'' in terms of massive momenta at $\log\omega$ and $\omega(\log\omega)^2$ level \cite{Sahoo:2021ctw}. We also present their explicit small-deflection expansions, as well as a conjecture for the leading logarithmic terms $\omega^{n-1}(\log\omega)^{n}$ for any $n\ge3$, that we later cross-check at Newtonian level in the case of a $2\to2$ background process.  Section~\ref{enspect} is then devoted to the analysis of the low-frequency energy emission spectrum, first for generic kinematics and then in the PM expansion. This section contains in particular the comparison with the ultrarelativistic limit and with the PN results mentioned above. We conclude in Section~\ref{sec:timedomain} with a  discussion of the waveform in time domain. The leading contribution in the PN limit provides a convenient way to generate simple cross-checks for the previous results, including in particular a closed-form expression for the Newtonian quadrupole that captures all leading logarithms. We also give the time-domain counterpart of the resummed formula capturing all leading logarithms for a $2\to2$ hard process for generic velocities. The paper also contains several appendices. Appendix~\ref{app:integrals} provides some details on the evaluation of the angular integrals involved in the energy spectrum, Appendix~\ref{app:FTtomega} deals with Fourier transforms from time to frequency domain, while Appendix~\ref{AppC} and Appendix~\ref{app:newt} contain more details on the derivations performed in Subsections~\ref{PMwaveform} and \ref{NewtonianLimit}.

\section{Low-frequency waveform}
\label{sec:lowfreq} 

Refs.~\cite{Sahoo:2018lxl,Saha:2019tub,Sahoo:2020ryf,Sahoo:2021ctw} studied in detail the low-frequency expansion of the spectral waveform $\tilde{w}^{\mu\nu}$, which is defined by
\begin{equation}\label{eq:conventiongmeta}
    g_{\mu\nu} - \eta_{\mu\nu} \sim \frac{4G}{r}\,w_{\mu\nu}(t, n)\,,
    \qquad
    \tilde{w}_{\mu\nu}(\omega n) = \int_{-\infty}^{+\infty} e^{i\omega t}\, {w}_{\mu\nu}(t, n)\,dt\,,
\end{equation}
for large $r$, fixed retarded time $t$ and angles characterized by the null vector $k^\mu = \omega\,n^\mu$. It takes the following form\footnote{The small-frequency expansion also involves terms of the type $\omega^{n-1}(\log\omega)^{m}$ with $0\le m\le n-1$, such as $\omega^0$ and $\omega\log\omega$, which are absorbed by the ellipses in Eq.~\eqref{eq:wtildesoft}.}
in the limit of small frequency $\omega$,
\begin{equation}\label{eq:wtildesoft}
    \tilde{w}^{\mu\nu}
    \sim
    \frac{i}{\omega}\,A_0^{\mu\nu}
    - \sum_{n=1}^\infty (-i\omega)^{n-1}\frac{(\log\omega)^n}{n!}\,A_n^{\mu\nu}
     + \cdots
     \qquad
    \text{as }\omega\to 0^+\,.
\end{equation} 
In particular, those references obtained general expressions for the three coefficients $A_0^{\mu\nu}$, $A_1^{\mu\nu}$, $A_2^{\mu\nu}$. The key idea is that soft gravitational-wave emissions can be obtained by suitably dressing a background hard process involving particles with momenta $k_a^\mu$. 

\subsection{Classical soft theorems}
\label{sec2.1}

We consider first the case in which this background process can include massive and massless particles.
Letting
\begin{equation}
    K_\text{in/out}^\mu=
    \sum_{a\in\text{in}/\text{out}} k_a^\mu\,,\qquad
    E_{\text{in}/\text{out}} = -K_\text{in/out}\cdot n\,,
\end{equation}
we have in this case
\begin{equation}
    K_\text{in}^\mu = K_\text{out}^\mu \,,
    \qquad
    E_\text{out} = E_\text{in} \equiv E
\end{equation}
by total momentum conservation. 
Note that $E$ coincides with the center-of-mass energy $E = \sqrt{s}$ if one decomposes $n^\mu = (1,\hat n)$ in that frame, while for generic decompositions $E\neq \sqrt{s}$.
It is also convenient to define 
\begin{equation}
p_a^\mu = \eta_a k_a^\mu
\end{equation}
where $\eta_a =-1$ if $a$ corresponds to an incoming state and $\eta_a=+1$ if $a$ is outgoing.

We denote by $\tau_{ab}$ the relative logarithmic drift of the particle worldlines, 
\begin{equation}
    \tau_{ab} = - \frac{d f(\zeta)}{d\zeta}\Big|_{\zeta = -p_a\cdot p_b/(m_a m_b)}\,,
\end{equation}
where the function $f$ is fixed by the exponentiation of phase infrared divergences due to massless exchanges between hard external lines \cite{Krishna:2023fxg,Alessio:2024wmz}. We will focus on minimally coupled massive scalars in general relativity, for which (focusing on $a\neq b$ and $\eta_a=\eta_b$)
\begin{equation}\label{eq:tauGR}
f_{\mathrm{GR}}(\zeta)=\frac{\zeta^2-\frac{1}{2}}{\sqrt{\zeta^2-1}}\,,
\qquad
\tau_{ab}
=
    p_a\cdot p_b\, \frac{(p_a\cdot p_b)^2-\tfrac{3}{2}p_a^2 p_b^2}{\left[(p_a\cdot p_b)^2-p_a^2 p_b^2\right]^{\tfrac{3}{2}}}
= 
- \frac{\sigma_{ab}(2\sigma_{ab}^2 -3)}{2(\sigma_{ab}^2-1)^{3/2}}\,.
\end{equation}
where we introduced the following symbol,
\begin{equation}\label{eq:sigmadef}
\sigma_{ab}= - \eta_a \eta_b\, \frac{p_a \cdot p_b}{m_a m_b}\,.
\end{equation}
Conversely, for $\mathcal{N}=8$ scalars that acquire masses via Kaluza--Klein compactification along orthogonal directions (see \cite{Caron-Huot:2018ape,Parra-Martinez:2020dzs,DiVecchia:2021bdo,Bjerrum-Bohr:2021vuf,DiVecchia:2023frv}), and can exchange dilatons in addition to gravitons, one would have (focusing again on $a\neq b$ and $\eta_a=\eta_b$)
\begin{equation}\label{eq:tauN8}
f_{\mathcal{N}=8}(\zeta)=\frac{\zeta^2}{\sqrt{\zeta^2-1}}\,,
\qquad
\tau_{ab}
=
    p_a\cdot p_b\, \frac{(p_a\cdot p_b)^2-2p_a^2 p_b^2}{\left[(p_a\cdot p_b)^2-p_a^2 p_b^2\right]^{\tfrac{3}{2}}}
=
-
\frac{\sigma_{ab}(\sigma_{ab}^2 -2)}{(\sigma_{ab}^2-1)^{3/2}}
\,.
\end{equation}
For later convenience we also introduce the symbol
\begin{equation}
\tau^{(\eta)}_{ab} = |\eta_a+\eta_b| \,\tau_{ab}\,,
\end{equation}
which by construction equals $2\tau_{ab}$ when $a$, $b$ are either both incoming or both ingoing, and vanishes otherwise.
Then \cite{Sahoo:2018lxl,Saha:2019tub,Sahoo:2020ryf,Sahoo:2021ctw},\footnote{Here and in the following, summation terms for $a=b$ or $a=c$ are dropped, as suggested by the antisymmetric nature of $p_{[b} p_{a]}$ and $p_{[c} p_{a]}$.}
\begin{subequations}\label{eq:A0A1A2}
\begin{align}
\label{eq:A0}
    A_0^{\mu\nu}
    &=
    -\sum_{a}\frac{p_a^\mu p_a^\nu}{p_a\cdot n}\,,
    \\
\label{eq:A1}
    A_1^{\mu\nu}
    &=
    (-2GE)
    \sum_{a}\frac{p_a^\mu p_a^\nu}{p_a\cdot n}
    +G \sum_{a,b}
    \frac{\tau^{(\eta)}_{ab} p_a^{\mu}}{p_a\cdot n} 
    n_\rho p_{[b}^\rho p_{a]}^{\nu}
    \,,\\
\label{eq:A2}
    \begin{split}
    A_2^{\mu\nu}
    &=
    (2G E)^2\sum_{a}\frac{p_a^\mu p_a^\nu}{p_a\cdot n}
    -2(2GE)
    G \sum_{a,b}
    \frac{\tau^{(\eta)}_{ab} p_a^{\mu}}{p_a\cdot n} 
    n_\rho p_{[b}^\rho p_{a]}^{\nu}+   
    G^2 \sum_{a,b,c}
    \frac{\tau^{(\eta)}_{ab}\tau^{(\eta)}_{ac}}{p_a\cdot n} 
    n_\rho p_{[b}^\rho p_{a]}^{\mu}
    n_\sigma p_{[c}^\sigma p_{a]}^{\nu}\,,
    \end{split}
\end{align}
\end{subequations}
where we adopt the convention $\ell_{[a}m_{b]} =\ell_{a}m_{b} -\ell_{b}m_{a}$ without additional factors, so that e.~g.
\begin{equation}
    \sum_{a,b}
    \frac{\tau^{(\eta)}_{ab} p_a^{\mu}}{p_a\cdot n} 
    n_\rho p_{[b}^\rho p_{a]}^{\nu}
    =
    \sum_{a,b}
    \frac{\tau^{(\eta)}_{ab} p_a^{\mu}p_a^\nu}{p_a\cdot n} 
    \,(n_\rho p_{b}^\rho) 
    -
    \sum_{a,b}
    \tau^{(\eta)}_{ab} p_a^{\mu}p_{b}^{\nu}
\end{equation}
(this equation also makes $A_1^{\mu\nu}$ and $A_2^{\mu\nu}$ manifestly symmetric).

The first few terms in each line of \eqref{eq:A0A1A2} suggest the exponentiation of a phase factor $\omega^{2iGE\omega}$, which can be in fact understood on general grounds as it is fixed by the phase infrared divergences associated to rescattering of the soft particle
\cite{Sahoo:2020ryf,Krishna:2023fxg,Alessio:2024wmz}. In particular, following \cite{Krishna:2023fxg}, such factor corresponds to the exponentiation of 
\begin{equation}
\label{Wph}
-\mathcal{W}_{\mathrm{phase}}=2iEG\omega\log\omega 
\end{equation}
(see Eq.~\eqref{resummedWeinb} below)
and it takes into account the infrared divergences coming from loop diagrams where a graviton is exchanged between the soft (massless) external line and a hard external line.
Up to $\omega(\log\omega)^2$ accuracy, \eqref{eq:A0A1A2} can be thus recast in the equivalent form 
\begin{equation}\label{eq:wtildesoftresummed}
    \tilde{w}^{\mu\nu}
    =
    e^{2iGE\omega \log\omega}
    \left[
    \frac{i}{\omega}\,a_0^{\mu\nu}
    - \sum_{n=1}^\infty (-i\omega)^{n-1}\frac{(\log\omega)^n}{n!}\,a_n^{\mu\nu}
     + \cdots
    \right]
\end{equation}
with
\begin{subequations}\label{eq:A0A1A2resummed}
\begin{align}
\label{softleading}
    a_0^{\mu\nu}
    &=
    -\sum_{a}\frac{p_a^\mu p_a^\nu}{p_a\cdot n}
    =
    A_0^{\mu\nu}\,,
    \\
    \label{softsubleading}
    a_1^{\mu\nu}
    &=
    G \sum_{a,b}
    \frac{\tau^{(\eta)}_{ab} p_a^{\mu}}{p_a\cdot n} 
    n_\rho p_{[b}^\rho p_{a]}^{\nu}
    \,,
    \\\label{softsubsubxleading}
    a_2^{\mu\nu}
    &=G^2 \sum_{a,b,c}
    \frac{\tau^{(\eta)}_{ab}\tau^{(\eta)}_{ac}}{p_a\cdot n} 
    n_\rho p_{[b}^\rho p_{a]}^{\mu}
    n_\sigma p_{[c}^\sigma p_{a]}^{\nu}\,.    
\end{align}
\end{subequations}
Gauge invariance is manifestly satisfied by $a_0^{\mu\nu}n_\nu = 0$, which holds thanks to momentum conservation $\sum_a p_a^\mu =0$, and by $a_1^{\mu\nu}n_\nu = 0$, $a_2^{\mu\nu}n_\nu = 0$, which hold thanks to the identity $p_{[a}^{\mu}p_{b]}^{\nu} n_\mu n_\nu=0$ (similar considerations hold for $A_0^{\mu\nu}$, $A_1^{\mu\nu}$, $A_2^{\mu\nu}$).
Note that, in the PM expansion, $a_0^{\mu\nu}$, $a_1^{\mu\nu}$ and $a_2^{\mu\nu}$ start at order $\mathcal{O}(G)$, $\mathcal{O}(G)$ and $\mathcal{O}(G^3)$ respectively, in agreement with \cite{Georgoudis:2023eke,Alessio:2024wmz}. This is because the relative sign leads to a cancellation between incoming and outgoing states in $a_0^{\mu\nu}$ and $a_2^{\mu\nu}$ for small deflections, while $a_1^{\mu\nu}$ has a nontrivial $Q\to0$ limit.
The relations between $A_\ell^{\mu\nu}$ and $a_\ell^{\mu\nu}$ for general $\ell\ge1$ read
\begin{subequations}
\begin{align}
\label{eq:linkAlal}
A_{\ell}^{\mu\nu}&=-(-2GE)^{\ell}a_0^{\mu\nu}+\sum_{n=1}^{\ell}(-2GE)^{\ell-n}\binom{\ell}{n}a_n^{\mu\nu}\,,
\\
a_{\ell}^{\mu\nu}&=-(2GE)^{\ell}A_0^{\mu\nu}+\sum_{n=1}^{\ell}(2GE)^{\ell-n}\binom{\ell}{n}A_n^{\mu\nu}\,.
\end{align}
\end{subequations}

The coefficients
$a_0^{\mu\nu}$, $a_1^{\mu\nu}$, $a_2^{\mu\nu}$ can be also obtained from the soft operator
\begin{equation}\label{eq:softoperator}
\hat{\mathcal{S}}^{\mu\nu}
=
\sum_{a}
\left[\frac{p_a^{\mu}p_a^{\nu}}{p_a\cdot k}
+
\frac{p_a^{(\mu}k_{\rho}\hat{J}^{\nu)\rho}_a}{p_a\cdot k}
+\frac{1}{2}\frac{k_{\rho}\hat{J}^{\mu\rho}_a k_{\sigma}\hat{J}^{\nu\sigma}_a}{p_a\cdot k}
+\cdots
\right],
\end{equation}
where we adopt the convention $\ell_{(\mu}m_{\nu)}=\frac{1}{2}(\ell_\mu m_\nu+\ell_\nu m_\mu)$, the $\cdots$ correspond to terms scaling as $\omega^n$ with $n>1$ and we define
\begin{equation}
\hat{J}_a^{\mu\nu} = p_a^{\mu} \frac{\partial}{\partial p_{a\nu}} - p_a^{\nu} \frac{\partial}{\partial p_{a\mu}}\,.
\end{equation}
Indeed, letting \cite{Krishna:2023fxg}
\begin{align}
\label{Wreg}
\mathcal{W}_{\mathrm{reg}}= - \frac{iG}{2}\log\omega \sum_{a\neq b}
|\eta_a+\eta_b| f_{\mathrm{GR}}(\sigma_{ab})m_am_b\,,
\end{align}
where $f_{\mathrm{GR}}$ is given in \eqref{eq:tauGR}, 
equation \eqref{eq:wtildesoftresummed} can be obtained by dressing the soft operator \eqref{eq:softoperator} according to
\begin{align}
\label{resummedWeinb}
\tilde{w}^{\mu\nu}=-i e^{-\mathcal{W}_{\mathrm{phase}}}e^{-\mathcal{W}_{\mathrm{reg}}}\hat{\mathcal{S}}^{\mu\nu}e^{\mathcal{W}_{\mathrm{reg}}}\,.
\end{align}
In particular, $\frac{1}{\omega}$, $\log\omega$ and $\omega(\log\omega)^2$ are given by the terms in which all derivatives in $\hat{\mathcal{S}}^{\mu\nu}$ act on ${\mathcal{W}}_\text{reg}$ appearing in the exponent\footnote{This would correspond to discard terms in \eqref{resummedWeinb} having more than two derivatives acting on $\mathcal{W}_{\mathrm{reg}}$ which only yield subleading logarithms. In fact, such terms are also subleading in the classical limit, as can be seen by restoring $\hbar$ in the various expressions.} via 
\begin{equation}
k_\rho \hat{J}_a^{\nu\rho}\mathcal{W}_\text{reg} = - i G \sum_b \tau_{ab}^{(\eta)}k_\rho p_{[b}^{\rho}p_{a]}^{\nu}\,\log\omega\,.
\end{equation}
Equation \eqref{resummedWeinb} shows explicitly that the classical soft graviton theorems can be directly obtained from the exponentiation of infrared divergences \cite{Weinberg:1965nx}, where one is performing the substitution $\epsilon^{-1}\mapsto\ 2\log\omega$ as also discussed in \cite{Agrawal:2023zea,Alessio:2024wmz}.

\subsection{Classical soft theorems rewritten}
\label{ssec:rewritten}

While \eqref{eq:A0A1A2} and \eqref{eq:A0A1A2resummed} are in principle completely general, their application requires prior knowledge of all initial and final momenta involved in the background hard scattering process. Since any gravitational collision of massive objects entails also the emission of gravitons with momenta $k^\mu$ and a certain phase-space distribution, the spectral emission rate $\rho(k)$
given by
\begin{equation}
\rho(k) = 8 \pi G\, \tilde{w}^{\mu\nu\ast}(k)
\left(\eta_{\mu\rho}\eta_{\nu\sigma}-\tfrac{1}{D-2}\,\eta_{\mu\nu}\eta_{\rho\sigma}\right) 
\tilde{w}^{\rho\sigma}(k)\,,
\end{equation}
for practical applications it is more convenient to regard as background process only the one involving the massive particles. 

This allows for a rewriting \cite{Sahoo:2021ctw} whereby one isolates the contributions due to final hard gravitons\footnote{In addition, similar contributions associated to emissions of dilatons and Kaluza--Klein scalars and vectors should be taken into account in $\mathcal{N}=8$ supergravity.} 
and restricts all sums to \emph{massive} hard states only, 
\begin{equation}\label{eq:replace}
    \sum_{a\in\text{out}} \mapsto 
    \sum_{a\in\text{out}} + \int_k \rho(k)
\end{equation}
where we adopt $\int_k$ as a shorthand for the integral over the phase space of gravitons with momentum $k^\mu = (|\vec k|, \vec k\,)=\omega\, n^\mu$,
\begin{align}
\int_k=\int 2\pi\theta(k^0)\,\delta(k^2)\,\frac{d^D k}{(2\pi)^D}=\int\frac{d^{D-1}\vec{k}}{2|\vec{k}|(2\pi)^{D-1}}
=
    \int_0^\infty \omega^{D-2}\, d\omega \int \frac{d\Omega(n)}{2\omega(2\pi)^{D-1}}\,.
\end{align}
In this way,\footnote{$a_\text{NL}^{\mu\nu}$ exhibits a collinear divergence in $D=4$, which however disappears when considering the projection over physical polarizations, or when contracting with $n_\nu$ since $a_\text{NL}^{\mu\nu}n_\nu = - \boldsymbol{P}^\mu$.}
\begin{equation}\label{eq:nonlinearmemory}
    A_0^{\mu\nu}
    =
    A_0^{(+)\mu\nu}
    -
    A_0^{(-)\mu\nu}
    +
    a^{\mu\nu}_\text{NL}\,,\qquad
    a^{\mu\nu}_\text{NL}
    =
    -\int_k \rho(k)\, \frac{k^\mu k^\nu}{k\cdot n}
    \,,
\end{equation}
where
\begin{equation}\label{eq:A0pm}
    A_0^{(+)\mu\nu}
    =
    -\sum_{a\in\text{out}}\frac{p_a^\mu p_a^\nu}{p_a\cdot n}\,,
    \qquad
    A_0^{(-)\mu\nu}
    =
    \sum_{a\in\text{in}}\frac{p_a^\mu p_a^\nu}{p_a\cdot n}
\end{equation}
encode the linear memory effect,
while $a_\text{NL}^{\mu\nu}$ is the nonlinear memory contribution \cite{Christodoulou:1991cr,Wiseman:1991ss,Thorne:1992sdb}. 
One can rewrite the latter in terms of the emitted energy of the radiation as a function of the solid angle,
\begin{align}
\frac{dE}{d\Omega}(n)=\int_0^{\infty}d\omega \,\omega^{D-2}\,\frac{\rho(\omega\, n)}{2(2\pi)^{D-1}}\,,
\end{align}
according to \cite{Christodoulou:1991cr,Wiseman:1991ss,Thorne:1992sdb}
\begin{equation}
a_\text{NL}^{\mu \nu} =- \int d\Omega(n')\frac{dE}{d \Omega}(n') \, \frac{n'^\mu n'^\nu}{n'\cdot n}
    \label{aNLnew}
\end{equation}
where the integral is performed over the solid angle surrounding the source.

We thus start from the expressions \eqref{eq:A0A1A2} and apply the rewriting \eqref{eq:replace}. Let us note that
\begin{equation}
\label{radmomP}
    \boldsymbol{P}^\mu = \int_k \rho(k)\,k^\mu\,,\qquad
    \sum_a p_a^\mu = K_\text{out}^\mu - K_\text{in}^\mu = - \boldsymbol{P}^\mu\,,
\end{equation}
so that now the total ``energy scale'' $E$ used in the previous section coincides with $E_\text{in}$, while
\begin{equation}
E_{\text{out}} = E_\text{in} + \boldsymbol{P}\cdot n\,.
\end{equation}
Using these relations and exploiting the fact that $\tau_{ab}=-1$ whenever either $a$ or $b$ is massless, one can show that
\begin{equation}
    A_1^{\mu\nu}
    =
    A_1^{(+)\mu\nu}
    -
    A_1^{(-)\mu\nu}
    \,,
    \qquad
    A_2^{\mu\nu}
    =
    A_2^{(+)\mu\nu}
    -
    A_2^{(-)\mu\nu}
\end{equation}
where
\begin{subequations}
\begin{align}
    A_1^{(+)\mu\nu}
    &=
    -2G 
    E_\text{out}\sum_{a\in\text{out}}\frac{p_a^\mu p_a^\nu}{p_a\cdot n}
    -2GK_\text{out}^\mu K_\text{out}^\nu
    + 2G \sum_{a,b\in\text{out}}
    \frac{\tau_{ab} p_a^{\mu}}{p_a\cdot n} 
    n_\rho p_{[b}^\rho p_{a]}^{\nu}\,,
    \\
    A_1^{(-)\mu\nu}
    &=
    2G 
    E_\text{in}\sum_{a\in\text{in}}\frac{p_a^\mu p_a^\nu}{p_a\cdot n}
    -2GK_\text{in}^\mu K_\text{in}^\nu
    - 2G \sum_{a,b\in\text{in}}
    \frac{\tau_{ab} p_a^{\mu}}{p_a\cdot n} 
    n_\rho p_{[b}^\rho p_{a]}^{\nu}\,,
\end{align}
\label{AAA}
\end{subequations}
and
\begin{subequations}
\begin{align}
\begin{split}
    A_2^{(+)\mu\nu}
    &=
    (2G E_\text{out})^2\sum_{a\in\text{out}}\frac{p_a^\mu p_a^\nu}{p_a\cdot n}
    +2G (2GE_\text{out})K_\text{out}^\mu K_\text{out}^\nu
    \\
    &-2 (2GE_{\text{out}})
    2G \sum_{a,b\in\text{out}}
    \frac{\tau_{ab} p_a^{\mu}}{p_a\cdot n} 
    n_\rho p_{[b}^\rho p_{a]}^{\nu}
    + (2G)^2 \sum_{a,b,c\in\text{out}}
    \frac{\tau_{ab} \tau_{ac}}{p_a\cdot n} 
    n_\rho p_{[b}^\rho p_{a]}^{\mu}
    n_\sigma p_{[c}^\rho p_{a]}^{\nu}\,,
    \end{split}
    \\
\begin{split}
    A_2^{(-)\mu\nu}
    &=
    -(2G E_{\text{in}})^2\sum_{a\in\text{in}}\frac{p_a^\mu p_a^\nu}{p_a\cdot n}
    +2G (2GE_\text{in})K_\text{in}^\mu K_\text{in}^\nu
    \\
    &+2 (2GE_{\text{in}})
    2G \sum_{a,b\in\text{in}}
    \frac{\tau_{ab} p_a^{\mu}}{p_a\cdot n} 
    n_\rho p_{[b}^\rho p_{a]}^{\nu}
    - (2G)^2 \sum_{a,b,c\in\text{in}}
    \frac{\tau_{ab} \tau_{ac}}{p_a\cdot n} 
    n_\rho p_{[b}^\rho p_{a]}^{\mu}
    n_\sigma p_{[c}^\rho p_{a]}^{\nu}\,.
    \end{split}
\end{align}
\label{BBB}
\end{subequations}
The remarkable feature of the above equations is that, \emph{unlike} $A_0^{\mu\nu}$, thanks to nontrivial cancellations among the terms in \eqref{eq:A1}, \eqref{eq:A2}, the coefficients $A_{1,2}^{\mu\nu}$ do not involve the detailed form of the spectral emission rate $\rho(k)$ that enters the nonlinear memory effect.
These cancellations can be traced back to the exponentiation of infrared divergences in gravity \cite{Weinberg:1965nx}. Indeed, the scattering amplitude of a general process $\alpha\to\beta$ involving $N+1$ massive particles will factorise as $\mathcal{A}_{\alpha\to\beta}=e^{\mathcal{W}}[\mathcal{A}_{\alpha\to\beta}]_{\mathrm{IR\,finite}}$ at any loop order, where the function $\mathcal{W}$ encoding all the infrared divergences is given by
\begin{align}
\label{Wfunction}
\mathcal{W}=\frac{G}{2\pi\epsilon}\sum_{a,b}
m_a m_b f_{\mathrm{GR}}(\sigma_{ab})\,\eta_a\eta_b\mathrm{arccosh} \sigma_{ab}
-
\frac{iG}{2\epsilon}\sum_{a\neq b}m_a m_b f_{\mathrm{GR}}(\sigma_{ab})\,\delta_{\eta_a,\eta_b}\,.
\end{align}
When considering the limit where one  particle is taken to be massless, $m_{N+1}\to 0$ and it has momentum $k^{\mu}=\omega\, n^{\mu}$, the function $\mathcal{W}$ is actually smooth (which is also the key property ensuring the absence of collinear divergences in gravity), and can be separated as
\begin{align}
\mathcal{W}=\mathcal{W}_{\mathrm{reg}}+\mathcal{W}_{\mathrm{phase}},
\end{align}
that, after the classical limit (which amounts to a near-forward limit and leads to discarding the term having $\mathrm{arccosh\sigma_{ab}}$ in \eqref{Wfunction})\footnote{For a detailed explanation of the classical limit see Appendix~B of~\cite{Alessio:2024wmz}.} and the substitution ${\epsilon}^{-1}\mapsto 2\log\omega$, match $\mathcal{W}_{\mathrm{reg}}$ and $\mathcal{W}_{\mathrm{phase}}$ in \eqref{Wreg} and \eqref{Wph}.

Naturally, gauge invariance is still upheld, but now it relies on the balance law \eqref{radmomP}, in particular linear $A_0^{(+)\mu\nu}-A_0^{(-)\mu\nu}$ and nonlinear $a_\text{NL}^{\mu\nu}$ memory contributions are not separately gauge invariant, as only their sum obeys $A_0^{\mu\nu}n_\nu=0$.

Finally, we find the following resummed form,
\begin{align}\label{eq:wtildesoftresummedNL}
\begin{split}
    \tilde{w}^{\mu\nu}
    &=
    e^{2iGE_\text{out}\omega\log\omega}
    \left[
    \frac{i}{\omega}\,a_0^{(+)\mu\nu}
    - \sum_{n=1}^\infty (-i\omega)^{n-1}\frac{(\log\omega)^n}{n!}\,a_n^{(+)\mu\nu}
    +\cdots
    \right]
    \\
    &
    -
    e^{2iGE_\text{in}\omega\log\omega}
    \left[
    \frac{i}{\omega}\,a_0^{(-)\mu\nu}
    - \sum_{n=1}^\infty (-i\omega)^{n-1}\frac{(\log\omega)^n}{n!}\,a_n^{(-)\mu\nu}
    +\cdots
    \right]
    \\
    &+\frac{i}{\omega}\left(
    a_\text{NL}^{\mu\nu}  
    - \frac{K_\text{out}^\mu K_\text{out}^\nu}{K_\text{out}\cdot n}
    + \frac{K_\text{in}^\mu K_\text{in}^\nu}{K_\text{in}\cdot n}
    \right)
\end{split}
\end{align}
with 
\begin{subequations}\label{eq:A0A1A2resummedNL}
\begin{align}
    a_0^{(+)\mu\nu}
    &=
    - \sum_{a\in\text{out}}\frac{p_a^\mu p_a^\nu}{p_a\cdot n}
    + \frac{K_\text{out}^\mu K_\text{out}^\nu}{K_\text{out}\cdot n}\,,
    \\
    a_1^{(+)\mu\nu}
    &=
    2G \sum_{a,b\in\text{out}}
    \frac{\tau_{ab} p_a^{\mu}}{p_a\cdot n} 
    n_\rho p_{[b}^\rho p_{a]}^{\nu}
    \,,
    \\
    a_2^{(+)\mu\nu}
    &= (2G)^2 \sum_{a,b,c\in\text{out}}
    \frac{\tau_{ab} \tau_{ac}}{p_a\cdot n} 
    n_\rho p_{[b}^\rho p_{a]}^{\mu}
    n_\sigma p_{[c}^\rho p_{a]}^{\nu}\,,    
\end{align}
\end{subequations}
while
\begin{subequations}\label{eq:A0A1A2resummedNLminus}
\begin{align}
    a_0^{(-)\mu\nu}
    &=
    \sum_{a\in\text{in}}\frac{p_a^\mu p_a^\nu}{p_a\cdot n}
    + \frac{K_\text{in}^\mu K_\text{in}^\nu}{K_\text{in}\cdot n}\,,
    \\
    a_1^{(-)\mu\nu}
    &=
    -
    2G \sum_{a,b\in\text{in}}
    \frac{\tau_{ab} p_a^{\mu}}{p_a\cdot n} 
    n_\rho p_{[b}^\rho p_{a]}^{\nu}
    \,,
    \\
    a_2^{(-)\mu\nu}
    &= -(2G)^2 \sum_{a,b,c\in\text{in}}
    \frac{\tau_{ab} \tau_{ac}}{p_a\cdot n} 
    n_\rho p_{[b}^\rho p_{a]}^{\mu}
    n_\sigma p_{[c}^\rho p_{a]}^{\nu}\,.   
\end{align}
\end{subequations}
We have included the terms with the initial and final total momenta $K_\text{in}^\mu$ and $K_\text{out}^\mu$ in
$a_0^{(\pm) \mu \nu}$ in order to correctly reproduce the terms with  the initial and final total momenta present in  \eqref{AAA} and \eqref{BBB} when we use  \eqref{eq:linkAlal}  for the components $\pm$, respectively.\footnote{We thank Biswajit Sahoo for a discussion on this point.} In this way, we are also consistent  with Eqs. (5.13) of~\cite{Sen:2024bax}. 

\subsection{PM expansion}
\label{PMwaveform}

The soft waveform in \eqref{eq:A0A1A2resummedNL} depends on both the initial and final momenta through the soft factors and it is therefore valid for any kinematic regime. Here we are interested in the PM expansion\footnote{In view of the overall factor of $4G/r$ in \eqref{eq:conventiongmeta}, an $\mathcal{O}(G^n)$ contribution into $\tilde{w}^{\mu\nu}$ yields an $\mathcal{O}(G^{n+1})$ contribution to the metric fluctuation $g_{\mu\nu}-\eta_{\mu\nu}$. Accordingly, $\tilde{w}_{n\text{PM}}^{\mu\nu}$ provides the $(n+1)$PM part of $g_{\mu\nu}-\eta_{\mu\nu}$. The metric fluctuation also contains $\mathcal{O}(G)$ static, $\delta(\omega)$, contributions which we do not consider here.} of $\tilde w^{\mu\nu}$
\begin{align}
\label{PMwf}
\tilde w^{\mu\nu}=\tilde w^{\mu\nu}_\text{1PM}+\tilde w^{\mu\nu}_\text{2PM}+\tilde w^{\mu\nu}_\text{3PM}+\cdots,
\end{align}
in the soft limit, where by $\tilde w^{\mu\nu}_{n\text{PM}}$ we denote the $n$PM contribution to the waveform corresponding to an $(n-1)$-loop amplitude calculation and scaling as $G^n$ in our conventions. Notice that from the expressions in \eqref{eq:wtildesoftresummed} in Section~\ref{sec2.1}, the soft factors $a_n^{\mu\nu}$ naively scale as $G^n$. However, the PM expansion in \eqref{PMwf} is achieved by relating the final momenta to the initial ones using the classical momentum exchanges,
\begin{align}
p_1^{\mu}+p_4^{\mu}=Q_1^{\mu},\qquad p_2^{\mu}+p_3^{\mu}=Q_2^{\mu},
\end{align}
and then by expanding the various soft factors in powers of $G$ using the perturbative expansion of $Q_1^{\mu}$ and $Q_2^\mu$. We define
\begin{align}
Q^{\mu}_{1}=Q^\mu+\boldsymbol{Q}^{\mu}_{1}+\mathcal{O}(G^4)\,,
\qquad
Q^{\mu}_{2}=-Q^\mu+\boldsymbol{Q}^{\mu}_{2}+\mathcal{O}(G^4)\,,
\end{align}
where\footnote{In our conventions, $b_e^\mu = b_1^\mu-b_2^\mu$ points toward particle 1.}
\begin{equation}
\label{PMmomenta1}
Q^\mu
=
-\frac{b_e^\mu}{b_e}\,Q\,,\qquad
Q
=
Q_{\mathrm{1PM}}
+
Q_{\mathrm{2PM}}
+
Q_{\mathrm{3PM}}
+
\mathcal{O}(G^4)\,,
\end{equation}
In this way, $Q_{n\mathrm{PM}}$ scales as $G^n$, while $\boldsymbol{Q}^{\mu}_1$ and $\boldsymbol{Q}^\mu_2$, which denote the radiative contributions, start at $\mathcal{O}(G^3)$ and satisfy
\begin{align}\label{eq:continuityBold}
\boldsymbol{Q}^{\mu}_1+\boldsymbol{Q}^{\mu}_2+\boldsymbol{P}^{\mu}=0\,,  
\end{align}
 where $\boldsymbol{P}^{\mu}$ is defined in \eqref{radmomP},
 so that the total energy-momentum conservation reads
 \begin{align}
Q_1^{\mu}+Q_2^{\mu}+\boldsymbol{P}^{\mu}=0\,.
 \end{align}
As already mentioned, using the above relations one can show that the PM expansions of $a_0^{\mu\nu}$, $a_1^{\mu\nu}$ and  $a_2^{\mu\nu}$ start at order $\mathcal{O}(G), \mathcal{O}(G)$ and $\mathcal{O}(G^3)$, respectively. Let us introduce the initial velocities $v_1^{\mu}$, $v_2^\mu$ of the particles as
\begin{align}
p_1^{\mu}=-m_1 v_{1}^{\mu},\qquad p_2^{\mu}=-m_1 v_{2}^{\mu},
\end{align}
such that $v^2_1=-1=v^2_2$. We are interested in the two-loop, 3PM  soft waveform for which we need the explicit expression for the classical exchanged momenta up to 3PM, which read
\cite{Bern:2019nnu,Bern:2019crd,Damour:2020tta,DiVecchia:2021ndb,Herrmann:2021tct}
\begin{subequations}\label{PMmomenta}
\begin{align}
Q_\text{1PM}
&=
\frac{2Gm_1m_2(2\sigma^2-1)}{b_e\sqrt{\sigma^2-1}} \,, 
\label{A1}\\
Q_\text{2PM} 
&=\frac{3\pi G^2m_1m_2(m_1+m_2)(5\sigma^2-1)}{4b_e^2\sqrt{\sigma^2-1}}\,,
\label{A2}
\\
Q_\text{3PM}
&=\frac{8 G^{3} m_{1}^{2} m_{2}^{2}}{b^{3}_e}\,\mathcal{F}(\sigma)\,,
\end{align}
\end{subequations}
and
\begin{equation}
\label{radmom}
\boldsymbol{Q}_{1}^{\mu}
=-\frac{G^3m_1^2m_2^2\mathcal{E}(\sigma)}{b_e^3}
\,\check{v}^{\mu}_{2}\,,
\qquad 
\boldsymbol{Q}_{2}^{\mu}
=-\frac{G^3m_1^2m_2^2\mathcal{E}(\sigma)}{b_e^3}
\,\check{v}^{\mu}_{1}\,,
\end{equation}
with
\begin{align}
\sigma=-v_1\cdot v_2\,,
\qquad\check{v}_1^{\mu}=\frac{\sigma v_2^{\mu}-v_1^{\mu}}{\sigma^2-1}\,,
\qquad \check{v}_2^{\mu}=\frac{\sigma v_1^{\mu}-v_2^{\mu}}{\sigma^2-1}\,,
\end{align}
satisfying $\check{v}_i\cdot v_j=-\delta_{ij}$ for $i,j=1,2$ and where 
\begin{align}
\mathcal{F}(\sigma)
&= \frac{\left(2 \sigma^{2}-1\right)^{2}\left(8-5 \sigma^{2}\right)}{6\left(\sigma^{2}-1\right)^{2}}
		-\frac{\sigma\left(14 \sigma^{2}+25\right)}{3 \sqrt{\sigma^{2}-1}} \\
		 	&+\frac{ 2s  (12 \sigma^4 - 10 \sigma^2 +1) }{4 m_{1} m_{2}\left(\sigma^{2}-1\right)^{\frac{5}{2}}}
				+\operatorname{arccosh}\sigma \left[
			\frac{\sigma\left(2 \sigma^{2}-1\right)^{2}\left(2 \sigma^{2}-3\right)}{2\left(\sigma^{2}-1\right)^{\frac{5}{2}}}
		+\frac{-4 \sigma^{4}+12 \sigma^{2}+3}{\sigma^{2}-1}\right],
  \nonumber
 \\ 
 \frac{\mathcal{E}(\sigma)}{\pi} 
 &= f_1(\sigma)+ f_2(\sigma) \log \frac{\sigma+1}{2} +f_3(\sigma) \frac{\sigma \operatorname{arccosh}\sigma}{2\sqrt{\sigma^2-1}}
\end{align}
with 
\begin{equation}\label{calE}
	\begin{split}
		f_1(\sigma) &= \frac{210 \sigma ^6-552 \sigma ^5+339 \sigma ^4-912 \sigma ^3+3148 \sigma ^2-3336 \sigma +1151}{48(\sigma^2-1)^{3/2}}\,,\\
		f_2(\sigma) &= -\frac{35 \sigma^4+60 \sigma^3-150 \sigma^2+76 \sigma -5}{8\sqrt{\sigma^2-1}}\,,\\
		f_3(\sigma) &= \frac{\left(2 \sigma^2-3\right) \left(35 \sigma^4-30 \sigma^2+11\right)}{8 \left(\sigma^2-1\right)^{3/2}}\,.
	\end{split}
\end{equation}
Note that $\mathcal{F}(\sigma)$ includes 3PM radiation-reaction and that, by \eqref{eq:continuityBold}, \eqref{radmom},
\begin{equation}\label{eq:radmomExpl}
\boldsymbol{P}^{\mu} =
\frac{G^3 m_1^2 m_2^2}{b_e^3}\,\mathcal{E}(\sigma)\,\frac{v_1^\mu+v_2^\mu}{\sigma+1}\,.
\end{equation}

Here, motivated by the simplicity of the expressions, we find it convenient to present the soft waveform in terms of the variables $\tilde{p}_1$, $\tilde{p}_2$, defined as
\begin{align}
\label{A}
\tilde{p}_1^{\mu}
=-p_1^{\mu}+\frac{Q_1^{\mu}}{2}
=p_4^{\mu}-\frac{Q_1^{\mu}}{2}
\equiv \tilde{m}_1\tilde{u}_1^{\mu},
\qquad 
\tilde{p}_2^{\mu}
=-p_2^{\mu}+\frac{Q_2^{\mu}}{2}
=p_3^{\mu}-\frac{Q_2^{\mu}}{2}
\equiv \tilde{m}_2\tilde{u}_2^{\mu},
\end{align}
which are orthogonal to the eikonal impact parameter $b_e^{\mu}$ appearing in the classical exchanged impulse above, $\tilde{p}_1\cdot b_e=0$, $\tilde{p}_2\cdot b_e=0$. 
Furthermore, we define the dimensionless quantities $\tilde{\alpha}_1$, $\tilde{\alpha}_2$
\begin{align}
\label{C}
\tilde{\alpha}_1=-\frac{\tilde{p}_1\cdot  k}{\tilde{m}_1\omega},\qquad \tilde{\alpha}_2=-\frac{\tilde{p}_2\cdot  k}{\tilde{m}_2\omega}\,.
\end{align}
Notice that the masses $m_1$, $m_2$ are related to  $\tilde{m}_1$, $\tilde{m}_2$ by
\begin{equation}
\label{masses}
\tilde{m}_1^2=m_1^2+\frac{Q^2}{4}\,,
\qquad
\tilde{m}_2^2=m_2^2+\frac{Q^2}{4}\,,
\end{equation}
which follow from $\tilde{u}_1^2=-1=\tilde{u}_2^2$,
while 
\begin{equation}\label{tildesigma}
\tilde{\sigma}
\equiv
-\tilde{u}_1\cdot \tilde{u}_2
=\frac{\sigma-\frac{Q^2}{4 m_1 m_2}}{\sqrt{1+\frac{Q^2}{4 m_1^2}}\sqrt{1+\frac{Q^2}{4m_2^2}}}
-
\frac{m_1+m_2}{m_1 m_2}\,\frac{G^3m_1^2m_2^2}{b_e^3}\,
\mathcal{E}(\sigma)
\,.
\end{equation}
We also recall \cite{Cristofoli:2021jas,DiVecchia:2023frv} that $b_e^\mu$ is the \emph{eikonal} impact parameter, which is related to the impact parameter $b^\mu$ orthogonal to the incoming momenta, $p_{1}\cdot b=0$, $p_{2}\cdot b=0$, by 
\begin{equation}
\label{bebJvec}
b_e^\mu = b^\mu - \left(
\frac{\check{v}_1^\mu}{2m_1}
-
\frac{\check{v}_2^\mu}{2m_2}
\right)
Q\,b_e\,,
\qquad 
Q = \frac{2m_1 m_2\sqrt{\sigma^2-1}\,\sin\frac{\Theta}{2}}{\sqrt{m_1^2+2 m_1 m_2 \sigma + m_2^2}}\,,
\end{equation}
where $\Theta$ denotes the deflection angle, so that
\begin{equation}
\label{bebJ}
b = b_e \cos\frac{\Theta}{2}\,.
\end{equation}
Note that the distinction between $m_1$, $m_2$, $\sigma$ and $\tilde m_1$, $\tilde m_2$, $\tilde \sigma$ starts being relevant at NNLO in the PM expansion, just like the difference in absolute value between $b$ and $b_e$ in \eqref{bebJ}. That is, in order to compare expressions written down using these two different sets of invariants, one needs to take into account feed-down terms in $\tilde{w}_{\text{3PM}}^{\mu\nu}$ that originate from the Taylor expansion of $\tilde{w}_{\text{1PM}}^{\mu\nu}$. This will be particularly important in order to compare with the PN-expanded expressions in Section~\ref{sec:timedomain}, where \eqref{bebJ}, \eqref{C},\eqref{masses}, \eqref{tildesigma} must be taken into account. 

We contract the free indices of the waveform with polarization vectors such that $\varepsilon\cdot n=0$, $\varepsilon\cdot \varepsilon=0$ and
define $\tilde w=\varepsilon_{\mu}\tilde{w}^{\mu\nu}\varepsilon_{\nu}$. Then, the soft expansion of the $n$PM waveform is
\begin{align}
\label{3PMsoft}
    \tilde{w}_{n\text{PM}}=\tilde w_{n\text{PM}}^{[\omega^{-1}]}\,\omega^{-1}+\tilde w_{n\text{PM}}^{[\log\omega]}\,\log\omega+\tilde w_{n\text{PM}}^{[\omega(\log\omega)^2]}\,\omega(\log\omega)^2+\cdots,
\end{align}
where by the ellipses we denote terms which are subleading in the soft expansion or are of the form $\omega^{n-1}(\log\omega)^{m}$ with $0\le m \le n-1$. We define the following structures,
\begin{subequations}\label{gstructures}
\begin{align}
\label{B1}
g_1
&=
[\tilde{\alpha}_1(\tilde{u}_2\cdot\varepsilon)-\tilde{\alpha}_2(\tilde{u}_1\cdot\varepsilon)]^2,
\\[10pt]
g_2
&=
\tilde{\alpha}_2^2 (\tilde{u}_1\cdot\varepsilon)^2 (\tilde{\alpha}_1 \sigma +\tilde{\alpha}_2)+\tilde{\alpha}_1^2 (\tilde{u}_2\cdot\varepsilon)^2 (\tilde{\alpha}_1+\tilde{\alpha}_2 \sigma )-2 \tilde{\alpha}_1 \tilde{\alpha}_2 (\tilde{u}_1\cdot\varepsilon) (\tilde{u}_2\cdot\varepsilon) (\tilde{\alpha}_1+\tilde{\alpha}_2),
\\[10pt]
g_3
&=
m_1^2 \tilde{\alpha}_1^4\,
[\tilde{\alpha}_2(b_e\cdot\varepsilon)+(b_e\cdot n)(\tilde{u}_2\cdot\varepsilon)]^2
-
m_2^2 \tilde{\alpha}_2^4\,
[\tilde{\alpha}_1(b_e\cdot\varepsilon)+(b_e\cdot n)(\tilde{u}_1\cdot\varepsilon)]^2,
\\[10pt]
g_4
&=
[\tilde{\alpha}_1(\tilde{u}_2\cdot \varepsilon)-\tilde{\alpha}_2(\tilde{u}_1\cdot \varepsilon)][\tilde{\alpha}_1(b_e\cdot n)(\tilde{u}_2\cdot\varepsilon)+\tilde{\alpha}_2(b_e\cdot n)(\tilde{u}_1\cdot\varepsilon)+2\tilde{\alpha}_1\tilde{\alpha}_2(b_e\cdot\varepsilon)],
\\[10pt]
\begin{split}
g_5
&=
(b_e\cdot n)^2[m_1\tilde{\alpha}_1^3(\tilde u_2\cdot\varepsilon)^2+m_2\tilde{\alpha}_2^3(\tilde u_1\cdot\varepsilon)^2]+2(b_e\cdot\varepsilon)(b_e\cdot n)\tilde{\alpha}_1 \tilde{\alpha}_2[m_1\tilde{\alpha}_1^2(\tilde u_2\cdot\varepsilon)\\
&+m_2\tilde{\alpha}_2^2(\tilde u_1\cdot\varepsilon)]+(b_e\cdot\varepsilon)^2\tilde{\alpha}_1^2\tilde{\alpha}_2^2(m_1\tilde{\alpha}_1+m_2\tilde{\alpha}_2)\,.\label{G}
\end{split}
\end{align}
\end{subequations}
Notice that also all the structures $g_1,\ldots,g_5$ in Eq. \eqref{gstructures} vanish when one substitutes one of the two polarizations vectors as $\varepsilon^{\mu}\mapsto n^{\mu}$,  except $g_2$, for which one finds
\begin{equation}\label{eq:g2nonginv}
g_2\big|_{\varepsilon\mapsto n}
=
-
\tilde{\alpha}_1^2
\tilde{\alpha}_2^2
(\sigma-1) (\tilde{u}_1+\tilde{u}_2)\cdot 
\varepsilon\,.
\end{equation}

In terms of  the structures $g_1,\ldots,g_5$ and of the explicit PM impulse given above, we find the following expressions for the soft expansion in the PM limit. At 1PM, in agreement with \cite{DiVecchia:2023frv,Georgoudis:2023eke,Alessio:2024wmz}, we have 
\begin{equation}\label{eq:treelevelwsummary}
\tilde{w}_{1\mathrm{PM}}^{[\omega^{-1}]}=\frac{i Q_\text{1PM}}{b_e\tilde{\alpha}_1^2\tilde{\alpha}_2^2}\,g_4
\,,
\qquad
\tilde{w}_{1\mathrm{PM}}^{[\log\omega]}
=
\frac{2G\tilde{m}_1\tilde{m}_2\sigma(2\sigma^2-3)}{\tilde{\alpha}_1\tilde{\alpha}_2(\sigma^2-1)^{\frac{3}{2}}}\,g_1
\,,
\qquad
\tilde{w}_{1\mathrm{PM}}^{[\omega(\log\omega)^2]}=0\,,
\end{equation}
while at 2PM, in agreement with \cite{Georgoudis:2023eke,Alessio:2024wmz},
we have (recalling $E=(p_1+p_2)\cdot n\simeq \tilde{m}_1\tilde{\alpha}_1+\tilde{m}_2\tilde{\alpha}_2$)
\begin{equation}\label{eq:oneloopwsummary}
\tilde{w}_{2\mathrm{PM}}^{[\omega^{-1}]}=\frac{iQ_{\text{2PM}}}{b_e\tilde{\alpha}_1^2\tilde{\alpha}_2^2}\,g_4\,,
\qquad
\tilde{w}_{2\mathrm{PM}}^{[\log\omega]}
=
2iGE\tilde{w}_{1\mathrm{PM}}^{[\omega^{-1}]}\,,
\qquad
\tilde{w}_{2\mathrm{PM}}^{[\omega(\log\omega)^2]}=2iGE \tilde{w}_{1\mathrm{PM}}^{[\log\omega]}\,.
\end{equation}
Finally, at 3PM, we obtain
\begin{subequations}
\begin{align}
\label{L}
\tilde{w}_\text{3PM}^{[\omega^{-1}]}&=\frac{i Q_{\text{3PM}}}{b_e\tilde{\alpha}_1^2\tilde{\alpha}_2^2}\,g_4
+\frac{i(b_e\cdot n)Q_{\text{1PM}}^3}{4b^3_e\tilde{m}_1^2\tilde{m}_2^2\tilde{\alpha}_1^4\tilde{\alpha}_2^4}\,g_3
-
\frac{iG^3\tilde{m}_1^2\tilde{m}_2^2\mathcal{E}(\sigma)}{b_e^3\tilde{\alpha}_1^2\tilde{\alpha}_2^2(\sigma^2-1)}\,g_2
+i\varepsilon\cdot a_{\mathrm{NL}}\cdot\varepsilon\,, 
\\[5pt]
\label{F}
\tilde{w}_\text{3PM}^{[\log\omega]}&=2i G E \tilde{w}_\text{2PM}^{[\omega^{-1}]}
+\frac{G E \sigma (2\sigma^2-3)Q_{\text{1PM}}^2}{2\tilde{m}_1\tilde{m}_2 b_e^2\tilde{\alpha}_1^3\tilde{\alpha}_2^3(\sigma^2-1)^{\frac{3}{2}}}\,g_5 \,,
\\[5pt]
\label{eq:omegalogomega23PM}
\tilde{w}_\text{3PM}^{[\omega(\log\omega)^2]}
&=-\frac{iG^2E^2[4(\sigma^2-1)^3 +\sigma^2 (2\sigma^2-3)^2]Q_{\text{1PM}}}{2b_e\tilde{\alpha}_1^2\tilde{\alpha}_2^2
(\sigma^2-1)^{3}}\,g_4\,.
\end{align}
\end{subequations}

Notice that $\tilde{w}^{[\log\omega]}$, $\tilde{w}^{[\omega(\log\omega)^2]}$, as well as $\tilde{w}_{1\text{PM}}^{[\omega^{-1}]}$, $\tilde{w}_{2\text{PM}}^{[\omega^{-1}]}$ are trivially gauge invariant since, when one substitutes one of the two polarizations vectors as $\varepsilon^{\mu}\mapsto n^{\mu}$, they all vanish thanks to the property mentioned below \eqref{gstructures},
\begin{align}
\label{ginv}
\tilde{w}^{[\log\omega]}\big|_{\varepsilon\mapsto n}=0\,,
\qquad 
\tilde{w}^{[\omega(\log\omega)^2]}\big|_{\varepsilon\mapsto n}=0\,, 
\qquad
\tilde{w}^{[\omega^{-1}]}_\text{1PM}\big|_{\varepsilon\mapsto n}=0\,,
\qquad
\tilde{w}^{[\omega^{-1}]}_\text{2PM}\big|_{\varepsilon\mapsto n}=0
\end{align}
whereas the $\omega^{-1}$ term at 3PM gives
\begin{align}
\label{leadinggaugeinv}
\tilde{w}_\text{3PM}^{[\omega^{-1}]}\big|_{\varepsilon\mapsto n}=i \left(n\cdot a_{\mathrm{NL}}\cdot\varepsilon
+
\frac{G^3m_1^2m_2^2[(\tilde u_1+\tilde u_2)\cdot\varepsilon]\mathcal{E}(\sigma)}{b^3_e(\sigma+1)}
\right),
\end{align}
and this vanishes thanks to
\begin{equation}
    n\cdot a_\text{NL} \cdot \varepsilon = - \boldsymbol{P}\cdot \varepsilon
    =
    -
    \frac{G^3m_1^2m_2^2[(\tilde u_1+\tilde u_2)\cdot\varepsilon]\mathcal{E}(\sigma)}{b^3_e(\sigma+1)}\,,
\end{equation}
where we used the expressions of $a_\text{NL}$ in \eqref{eq:nonlinearmemory} and of $\boldsymbol{P}^\mu$ in \eqref{radmomP}, \eqref{eq:radmomExpl}.
Therefore, as already noted in Subsection~\ref{ssec:rewritten}, at 3PM non-linear memory is needed in order to guarantee gauge invariance of the leading soft waveform and, as clear from \eqref{leadinggaugeinv}, this is related to the presence of radiative contributions to the classical momenta exchange encoded in the function $\mathcal{E}(\sigma)$ in \eqref{radmom}.
In other words, the splitting between linear and nonlinear memory is gauge-dependent and in principle ambiguous. However, $a_{\text{NL}}^{\mu\nu}$ starts at order $\mathcal{O}(p_\infty)$ in the small velocity limit, $p_\infty = \sqrt{\sigma^2-1}\to0$, \cite{Wiseman:1991ss,Favata:2011qi,Hait:2022ukn} which corresponds to a 2.5PN correction with respect to the leading $\mathcal{O}(p_\infty^{-4})$ Newtonian contribution to $\tilde{w}^{[\omega^{-1}]}_\text{3PM}$. We recall that one introduces a dimensionless frequency $u = \omega b/p_\infty$ which does not scale in the PN limit (see Section~\ref{sec:timedomain} for more details on the PN expansion). This means that, in the absence of an explicit PM expression for $a_\text{NL}$, \eqref{L} provides accurate expressions up to and including 2PN for $\tilde{w}_\text{3PM}^{[\omega^{-1}]}$. Instead, \eqref{F} and \eqref{eq:omegalogomega23PM} for $\tilde{w}_\text{3PM}^{[\log\omega]}$ and $\tilde{w}_\text{3PM}^{[\omega(\log\omega)^2]}$ are of course valid for generic velocities.

\subsection{A conjecture for \texorpdfstring{$\omega^{\ell-1}(\log\omega)^\ell$}{omega(ell-1)(log(omega)){ell}} and the resummed waveform}

The knowledge of the full $\mathcal{O}(\omega^2(\log\omega)^3)$ contribution to the 3PM waveform would require the corresponding soft factor, $a_3^{\mu\nu}$, which  has not been derived in general. However, inspired by Eq.~\eqref{resummedWeinb} and by the analysis of generalized soft operators in \cite{Li:2018gnc}\footnote{More precisely, in order to reach agreement with the PN expansion, the part that comes from the extended soft theorem of\cite{Li:2018gnc} must be supplemented by an additional gauge invariant term. We thank Biswajit Sahoo for a discussion on this point.} we conjecture that
\begin{align}\label{eq:a3c}
a_3^{\mu\nu}= G^3  \sum_{a,b,c,d}\frac{\tau^{(\eta)}_{ab} \tau^{(\eta)}_{ac} \tau^{(\eta)}_{ad}}{p_a\cdot n}\, n_{\rho}p^{\rho}_{[b}p^{\mu}_{a]}n_{\sigma}p^{\sigma}_{[c}p^{\nu}_{a]}n_{\alpha}(p_d^{\alpha}+p_a^{\alpha})\,.
\end{align}
Taking into account also the term coming from the exponentiation in \eqref{eq:wtildesoftresummed} via \eqref{eq:linkAlal}, this corresponds to the following expression for $A_3^{\mu\nu}$ which enters \eqref{eq:wtildesoft}, 
\begin{align}\label{eq:A3conjecture}
A_3^{\mu\nu}=(2GE)^3 a_{0}^{\mu\nu}+3(2GE)^2a_{1}^{\mu\nu}-3(2GE)a_2^{\mu\nu}+a_3^{\mu\nu},
\end{align}
in terms of $a_0^{\mu\nu}$, $a_1^{\mu\nu}$, $a_2^{\mu\nu}$ given in \eqref{eq:A0A1A2resummed} and $a_3^{\mu\nu}$  in \eqref{eq:a3c}. 
More explicitly, focusing on the $2\to2$ case and denoting 
\begin{subequations}
\begin{align}
\label{Bmunu}
B^{\mu \nu} (p_i)
&= \frac{p_2\cdot n}{p_1\cdot n} \, p_1^\mu p_1^\nu +
\frac{p_1\cdot n}{p_2\cdot n}\, p_2^\mu p_2^\nu - (p_1^ \mu p_2^\nu+p_1^\nu p_2^\mu)\,,\\[5pt]
B^{\mu \nu} (p_f)
&= \frac{p_3\cdot n}{p_4\cdot n}\, p_4^\mu p_4^\nu +\frac{p_4\cdot n}{p_3\cdot n}\, p_3^\mu p_3^\nu - (p_3^\mu p_4^\nu+p_3^\nu p_4^\mu)\,,
\end{align}
\end{subequations}
this translates to the following proposal for the first logarithm that is not fixed by the soft theorems of \cite{Sahoo:2018lxl,Saha:2019tub,Sahoo:2020ryf,Sahoo:2021ctw},
\begin{equation}
\label{GE4}
a_3^{\mu\nu} 
=- G^3\bigg[\bigg(\frac{\sigma_{12}(2\sigma_{12}^2-3)}{(\sigma^2_{12}-1)^{\frac{3}{2}}}\bigg)^3 (p_1\cdot n+p_2\cdot n)^{2}
B^{\mu\nu}(p_i)+(1,2)\leftrightarrow (3,4)\bigg].
\end{equation}
In particular, Eqs.~\eqref{GE4}, \eqref{eq:A3conjecture} agree with Eq.~\eqref{eq:A30PN} below which is derived independently at Newtonian level (note that only $a_3^{\mu\nu}$ contributes to $A_3^{\mu\nu}$ at leading PN order).

Recalling that $a_0^{\mu\nu}\sim\mathcal{O}(G)$, $a_1^{\mu\nu}\sim\mathcal{O}(G)$, $a_2^{\mu\nu}\sim\mathcal{O}(G^3)$, we note that at 3PM the $\mathcal{O}(\omega^2(\log\omega)^3)$ comes from the second and last terms in Eq.~\eqref{eq:A3conjecture} and it is explicitly given by
\begin{align}
\label{omega2log3}
\tilde{w}_{\text{3PM}}^{[\omega^2(\log\omega)^3]}=&-\frac{G^3m_1m_2(m_1\tilde{\alpha}_1+m_2\tilde{\alpha}_2)^2}{\tilde{\alpha}_1\tilde{\alpha}_2}\frac{\sigma(2\sigma^2-3)}{(\sigma^2-1)^{\frac{3}{2}}}\bigg[4+\frac{\sigma^2(2\sigma^2-3)^2}{3(\sigma^2-1)^{3}}\bigg]g_1,
\end{align}
which correctly reproduces the leading Newtonian order (see Section~\ref{NewtonianLimit} below) and provides an extension to relativistic velocities. 
Moreover, the first five orders in the small-velocity expansion of \eqref{omega2log3} as $p_\infty = \sqrt{\sigma^2-1}\to0$, corresponding to a relative 2PN accuracy, agree with the recent results in \cite{Bini:2024ijq}.

Let us conclude this section by presenting the natural generalization of \eqref{eq:a3c} for all leading logarithmic terms with $\ell\ge3$ as well,
\begin{equation}\label{eq:aell}
\begin{split}
a_\ell^{\mu\nu}
&
= G^\ell\sum_{a,a_1\dots a_{\ell}}\frac{\tau_{aa_1}^{(\eta)}\tau_{aa_2}^{(\eta)}\cdots \tau_{aa_\ell}^{(\eta)}}{p_a\cdot n}\, 
n_{\rho}p^{\rho}_{[a_1}p^{\mu}_{a]}
\,
n_{\sigma}p^{\sigma}_{[a_2}p^{\nu}_{a]}
\\
&\times
n_{\alpha_3}(p_{a_3}^{\alpha_3}+p_a^{\alpha_3})
\,
n_{\alpha_4}(p_{a_4}^{\alpha_4}+p_a^{\alpha_4})\cdots n_{\alpha_\ell}(p_{a_\ell}^{\alpha_\ell}+p_a^{\alpha_\ell})\,.
\end{split}
\end{equation}
One then gets in the $2\rightarrow 2$ case
\begin{align}\label{eq:proposalANYk}
a_\ell^{\mu\nu}=(-1)^\ell G^\ell\bigg[\bigg(\frac{\sigma_{12}(2\sigma_{12}^2-3)}{(\sigma^2_{12}-1)^{\frac{3}{2}}}\bigg)^\ell(p_1\cdot n+p_2\cdot n)^{\ell-1}B^{\mu\nu}(p_i)+(1,2)\leftrightarrow (3,4)\bigg],
\end{align}
which, for elastic $2\to2$ kinematics, holds for any $\ell\ge1$.
In Section~\ref{sec:timedomain}, we will provide a cross-check of the all-$\ell$ expression \eqref{eq:proposalANYk} at leading order in the post-Newtonian limit, by showing that it is equivalent to \eqref{eq:A2kA2kp1Conjecture} below. Substituting \eqref{eq:proposalANYk} into \eqref{eq:wtildesoftresummed}, we find that, for elastic $2\to2$ kinematics, the expression can be resummed as follows
\begin{equation}
\label{eq:finalproposal2to2prel}
\tilde{w}^{\mu\nu}
=
\frac{i}{\omega}
\,
\omega^{2iGE\omega}
\left[ 
a_0^{\mu\nu}
-
\left( 
\omega^{iGE \omega h(\sigma)}
-
1
\right)
\frac{B^{\mu\nu}(p_i)}{E}
+
\left( 
\omega^{-iGE \omega h(\sigma)}
-
1
\right)
\frac{B^{\mu\nu}(p_f)}{E}
\right]
+\cdots
\end{equation}
with 
$h(\sigma)=\frac{\sigma(2\sigma^2-3)}{(\sigma^2-1)^{{3}/{2}}}$
and a further cancellation that follows from $E= n\cdot (p_1+p_2)=-n\cdot (p_3+p_4)$ and  from $p_1^\mu+p_2^\mu+p_3^\mu+p_4^\mu=0$, 
\begin{equation}
\label{eq:finalproposal2to2}
\tilde{w}^{\mu\nu}
=
-
\frac{i}{E \omega}
\, \omega^{2iGE\omega}
\left[ 
\omega^{iGE \omega h(\sigma)}
B^{\mu\nu}(p_i)
-
\omega^{-iGE \omega h(\sigma)}
B^{\mu\nu}(p_f)
\right]
+\cdots
\end{equation}
The resummed $2\to2$ expression \eqref{eq:finalproposal2to2} has been also found to agree with the results obtained by calculating the waveform in the near-probe limit \cite{Fucito:2024wlg}.\footnote{We are particularly grateful to Rodolfo Russo for discussions and communications on this comparison.}

For a generic hard process, instead, the more general conjecture \eqref{eq:aell} leads to the following resummed expression for the leading logarithmic contributions,
\begin{equation}
\label{eq:finalproposalNtoM}
\tilde{w}=-\frac{i}{\omega}\,\omega^{2iGE\omega}
\sum_{a,b,c} \tau_{ab}^{(\eta)} \tau_{ac}^{(\eta)}
\left(
\frac{\mathcal{C}^{abc}_0}{\mathcal{T}^2_a}
+
\frac{\mathcal{C}^{abc}_1}{\mathcal{T}_a}(-i
G\omega\log\omega)
+
\frac{\mathcal{C}^{abc}_\infty}{\mathcal{T}_a^2}
\frac{\omega^{-iG\mathcal{T}_a \omega}}{p_a\cdot n}
\right)+\cdots
\end{equation}
where we introduced
\begin{subequations}
\begin{align}
\mathcal{T}_a &=\sum_e \tau_{ae}^{(\eta)} (p_a+p_e)\cdot n\,,\\
\mathcal{C}^{abc}_0 &= 2 (p_a\cdot \varepsilon) (p_b\cdot n) (p_a+p_c)\cdot \varepsilon  
+
p_a\cdot n\left[(p_a\cdot \varepsilon)^2-(p_b\cdot \varepsilon)(p_c\cdot\varepsilon) \right],
\\
\mathcal{C}^{abc}_1 &= (n\cdot p_{[b} p_{a]}\cdot \varepsilon)\, (p_a+p_c)\cdot \varepsilon\,,\\
\mathcal{C}^{abc}_\infty &= 
(n\cdot p_{[b} p_{a]}\cdot \varepsilon)\,
(n\cdot p_{[c} p_{a]}\cdot \varepsilon)\,.
\end{align}
\end{subequations}
Eq.~\eqref{eq:finalproposalNtoM} reduces to \eqref{eq:finalproposal2to2} when restricting to elastic $2\to2$ kinematics for the background hard process.
In the case of an $N\to N$ process in which one of the objects is much heavier than the remaining ones, the expression \eqref{eq:finalproposalNtoM} simplifies to a sum of $N-1$ copies of \eqref{eq:finalproposal2to2} in which only the light objects are dynamical, and thus also agrees with results obtained in the near-probe limit  in Ref.~\cite{Fucito:2024wlg}.

By analogy with \eqref{eq:wtildesoftresummedNL}, \eqref{eq:A0A1A2resummedNL}, \eqref{eq:A0A1A2resummedNLminus}, it is natural to guess that that the pattern continues also for $\ell\ge3$ and that $a_\ell^{(+)\mu\nu}$ (respectively $a_\ell^{(-)\mu\nu}$) are obtained from $a_\ell^{\mu\nu}$ ($-a_\ell^{\mu\nu}$) by restricting all sums to the out (in) hard states. In this way, the analogs of \eqref{eq:finalproposal2to2prel}, \eqref{eq:finalproposal2to2}
for a $2\to2$ process that now includes dissipative effects and radiation in the final state take the following form,
\begin{align}
\nonumber
    \tilde{w}^{\mu\nu}
    &=
    \frac{i}{\omega}\,
    \omega^{2iGE_\text{out}\omega}
    \left[
    a_0^{(+)\mu\nu}
    + \frac{1}{E_{\mathrm{out}}}(\omega^{-i G E_{\mathrm{out}}\omega h(\sigma_{34})}-1)B^{\mu\nu}(p_f)
    \right]
    \\
\nonumber
    &
    -
    \frac{i}{\omega}\,
    \omega^{2iGE_\text{in}\omega}
    \left[
    a_0^{(-)\mu\nu}
    + \frac{1}{E_{\mathrm{in}}}(\omega^{i G E_{\mathrm{in}}\omega h(\sigma_{12})}-1)B^{\mu\nu}(p_i)
    \right]
    \\
\label{resummedlogs}
    &
    +\frac{i}{\omega}\left(
    a_\text{NL}^{\mu\nu}
    - \frac{K_\text{out}^\mu K_\text{out}^\nu}{K_\text{out}\cdot n}
    + \frac{K_\text{in}^\mu K_\text{in}^\nu}{K_\text{in}\cdot n}
    \right)+\cdots
\end{align}
and thus
\begin{equation}
\label{resummedlogs2}
\begin{split}
    \tilde{w}^{\mu\nu}
    &=
    \frac{i \omega^{2iGE_\text{out}\omega}}{E_\text{out}\omega}\,
    \omega^{-i G E_{\mathrm{out}}\omega h(\sigma_{34})} 
    B^{\mu\nu}(p_f)
    -
    \frac{i \omega^{2iGE_\text{in}\omega}}{E_\text{in}\omega}\,
    \omega^{i G E_{\mathrm{in}}\omega h(\sigma_{12})} 
    B^{\mu\nu}(p_i)
    \\
    &
    +\frac{i}{\omega}
    \left(
    a_\text{NL}^{\mu\nu}
    - \frac{K_\text{out}^\mu K_\text{out}^\nu}{K_\text{out}\cdot n}
    + \frac{K_\text{in}^\mu K_\text{in}^\nu}{K_\text{in}\cdot n}
    \right)+\cdots
\end{split}
\end{equation}

\section{Energy spectrum in the soft limit}
\label{enspect}

The energy emission spectrum $\frac{dE}{d\omega}$ has an expansion of the following type for small frequencies,
\begin{equation}\label{eq:ZFexpandedSpectrum}
\frac{dE}{d\omega}
= 
\left(
\frac{dE}{d\omega}
\right)_\text{ZFL}
+
\left(
\frac{dE}{d\omega}
\right)_{\omega^2 (\log\omega)^2} \omega^2 (\log\tilde\omega)^2
+ 
\left(
\frac{dE}{d\omega}
\right)_{\omega^2 \log\omega} \omega^2 \log\omega
+
\mathcal O(\omega^2)\,,
\end{equation}
where $\left(
\frac{dE}{d\omega}
\right)_\text{ZFL}$ is the zero-frequency limit (ZFL) calculated in \cite{DiVecchia:2022nna} and we defined\footnote{The distinction between $\omega$ and $\tilde\omega$ is introduced in order to simplify the expression of $\left(
\frac{dE}{d\omega}
\right)_{\omega^2 \log\omega}$.} 
\begin{equation}
\tilde \omega = \frac{e^{\gamma_E}\omega b}{2\sqrt{\sigma^2-1}}\,.
\end{equation}

Substituting \eqref{eq:wtildesoftresummed} into
\begin{equation}\label{eq:todointegrals}
\frac{dE}{d\omega}
=
\frac{2G}{\pi} \int \frac{d\Omega}{4\pi}\,\omega^2 \tilde w^{\mu\nu} \tilde w^{\rho\sigma\ast}\left(
\eta_{\mu\rho}\eta_{\nu\sigma}
-\tfrac{1}{2}
\eta_{\mu\nu}\eta_{\rho\sigma}
\right)
\end{equation}
one obtains general expressions for the first two terms in \eqref{eq:ZFexpandedSpectrum},
\begin{equation}
    \left(
\frac{dE}{d\omega}
\right)_\text{ZFL}
  = \frac{2G}{\pi}  
  \int \frac{d\Omega}{4\pi}\, a_0^{\mu\nu}  a_0^{\rho\sigma}\left(
\eta_{\mu\rho}\eta_{\nu\sigma}
-\tfrac{1}{2}
\eta_{\mu\nu}\eta_{\rho\sigma}
\right)
\end{equation}
and
\begin{equation}
    \left(
\frac{dE}{d\omega}
\right)_{\omega^2(\log\omega)^2}
  = \frac{2G}{\pi}  
  \int \frac{d\Omega}{4\pi}\, 
  \left(a_1^{\mu\nu}  a_1^{\rho\sigma}
  +
  a_0^{\mu\nu} a_2^{\rho\sigma}\right)\left(
\eta_{\mu\rho}\eta_{\nu\sigma}
-\tfrac{1}{2}
\eta_{\mu\nu}\eta_{\rho\sigma}
\right)\,.
\end{equation}
Note that, owing to the manifest cancellation of the overall phase factor in \eqref{eq:wtildesoftresummed}, it is slightly simpler to use \eqref{eq:wtildesoftresummed}, \eqref{eq:A0A1A2resummed}, as opposed to the non-resummed expressions \eqref{eq:wtildesoft}, \eqref{eq:A0A1A2}.
Plugging in the explicit expressions \eqref{eq:A0A1A2resummed}, we obtain
\begin{equation}\label{eq:LOint}
    \left(
\frac{dE}{d\omega}
\right)_\text{ZFL}
  = \frac{2G}{\pi} 
  \sum_{a,b}
  \int \frac{d\Omega}{4\pi}\frac{(p_a\cdot p_b)^2-\tfrac{1}{2}\,p_a^2p_b^2}{(p_a\cdot n)(p_b\cdot n)}
\end{equation}
and
\begin{equation}
\begin{split}
   \left(
\frac{dE}{d\omega}
\right)_{\omega^2(\log\omega)^2}
&= \frac{2G^3}{\pi} 
  \sum_{a,b,c,d}
  \int \frac{d\Omega}{4\pi}
  \, 
  \left(
\eta_{\mu\rho}\eta_{\nu\sigma}
-\tfrac{1}{2}
\eta_{\mu\nu}\eta_{\rho\sigma}
\right)\\
&\times
\Bigg[
\frac{\tau^{(\eta)}_{ab}p_a^\mu}{p_a\cdot n}n_\alpha p^{\alpha}_{[b}p_{a]}^{\nu}
\frac{\tau^{(\eta)}_{cd}p_c^\rho}{p_c\cdot n}n_\beta p^{\beta}_{[d}p_{c]}^{\sigma}
-
\frac{\tau^{(\eta)}_{ab} \tau^{(\eta)}_{ac}}{p_a\cdot n} 
    n_\alpha p_{[b}^\alpha p_{a]}^{\mu}
    n_\beta p_{[c}^\beta p_{a]}^{\nu}
    \frac{p_d^\rho p_d^\sigma}{p_d\cdot n}
  \Bigg],
\end{split}
\end{equation}
which we may rewrite as
\begin{equation}\label{eq:NLOint}
\begin{split}
    \left(
\frac{dE}{d\omega}
\right)_{\omega^2(\log\omega)^2}
&= \frac{2G^3}{\pi} \,
  \sum_{a,b,c,d}
  \int \frac{d\Omega}{4\pi}
  \frac{n_\alpha n_\beta}{(p_a\cdot n)(p_c\cdot n)}
  \, 
  \left(
\eta_{\mu\rho}\eta_{\nu\sigma}
-\tfrac{1}{2}
\eta_{\mu\nu}\eta_{\rho\sigma}
\right)\\
&\times
\tau^{(\eta)}_{ab}
\Bigg[
\tau^{(\eta)}_{cd}
p_a^\mu p^{\alpha}_{[b}p_{a]}^{\nu}
p_c^\rho p^{\beta}_{[d}p_{c]}^{\sigma}
-
\tau^{(\eta)}_{ad}
    p_{[b}^\alpha p_{a]}^{\mu}
    p_{[d}^\beta p_{a]}^{\nu}
    p_c^\rho p_c^\sigma
  \Bigg].
\end{split}
\end{equation}
In Appendix~\ref{app:integrals} we discuss the evaluation of the integrals entering \eqref{eq:LOint} and \eqref{eq:NLOint}.

\subsection{ZFL}

The result for the ZFL is Lorentz-invariant and reads
\begin{equation}\label{eq:ZFL}
    \left(
\frac{dE}{d\omega}
\right)_\text{ZFL}
=
\frac{2G}{\pi}
\sum_{a,b} m_a m_b \left(\sigma_{ab}^2-\tfrac12\right)
\eta_a\eta_b \,\frac{\operatorname{arccosh}\sigma_{ab}}{\sqrt{\sigma^2_{ab}-1}}
\end{equation}
where $a,b$ run over all hard states, $\sigma_{ab}$ are given in \eqref{eq:sigmadef},
and $\eta_a =1$ for outgoing states, $\eta_a=-1$ for incoming states. Let us recall that, for elastic $2\to2$ kinematics, $\sigma_{34}=\sigma_{12}$, $\sigma_{24}=\sigma_{23}$ and
\begin{equation}\label{eq:sigma_explicit}
    \sigma_{12}=\sigma\,,\qquad
    \sigma_{13}=\sigma-\frac{Q^2}{2 m_1 m_2}\,,\qquad
    \sigma_{14}=1+\frac{Q^2}{2m_1^2}\,,\qquad
    \sigma_{23}=1+\frac{Q^2}{2m_2^2}\,.
\end{equation}
We can also introduce 
\begin{equation}
    x_{ab} = \sigma_{ab} - \sqrt{\sigma^2_{ab}-1}\,, \qquad
    \sigma_{ab} = \frac{1}{2}\left(x_{ab}+\frac{1}{x_{ab}}\right),
\end{equation}
where $0<x_{ab}\le 1$ for later convenience. For instance, the ZFL takes the following form in terms of these variables,
\begin{equation}\label{eq:ZFLx}
    \left(
\frac{dE}{d\omega}
\right)_\text{ZFL}
=
-
\frac{G}{\pi}
\sum_{a,b} m_a m_b \,
\eta_a\eta_b \,\frac{\left(1+x_{ab}^4\right)\log x_{ab}}{x_{ab}(1-x_{ab}^2)}\,.
\end{equation}
In the massless limit, the ZFL reads as follows \cite{Gruzinov:2014moa,Ciafaloni:2015vsa,Ciafaloni:2015xsr,Sahoo:2021ctw},
\begin{equation}\label{eq:ZFLUR}
     \left(
\frac{dE}{d\omega}
\right)_\text{ZFL}
=
        \frac{4G}{\pi}
\left[
Q^2 \log\left(\frac{s}{Q^2}-1\right)
-s \log\left(1-\frac{Q^2}{s}\right)
\right],
\end{equation}
while in the PM limit, expanding to leading order for small $Q$,
\begin{equation}\label{eq:ZFL_PM}
    \left(
\frac{dE}{d\omega}
\right)_\text{ZFL}
=
\frac{G Q^2}{\pi}\,\mathcal{I}(\sigma)
+
\mathcal O(G^5)\,,
\end{equation}
where $Q = Q_\text{1PM} + Q_\text{2PM} + \mathcal O(G^3)$ and 
\begin{equation}\label{eq:mathcalI}
    \mathcal{I}(\sigma)
    =
    \frac{2}{\sigma^2-1}
    \left[
    \frac{8-5\sigma^2}{3}
    +
    \frac{\sigma(2\sigma^2-3)\operatorname{arccosh}\sigma}{\sqrt{\sigma^2-1}}
    \right].
\end{equation}

By substituting instead the deflection of the massive particles accurate up to and including $\mathcal{O}(G^3)$, see Eq.~\eqref{PMmomenta}, then the general expression \eqref{eq:ZFL} allows one to obtain a prediction for the ZFL that is reliable up to and including $\mathcal O(G^5)$ in the PM limit,
\begin{equation}\label{eq:ZFLextended}
\begin{split}
    \left(
\frac{dE}{d\omega}
\right)_\text{ZFL}
&=
\frac{G Q^2}{\pi}\,\mathcal{I}(\sigma)
\\
&
-
\frac{G Q^4}{\pi m_1 m_2 (\sigma^2-1)^2}
\left[ 
\frac{3}{2}\frac{\operatorname{arccosh}\sigma}{\sqrt{\sigma^2-1}}
+
\frac{\sigma}{2}(2\sigma^2-5)
+
\frac{2}{5}
\frac{m_1^2+m_2^2}{m_1 m_2}
(\sigma^2-1)^2
\right]
\\
&+
\mathcal O(G^6)\,.
\end{split}
\end{equation}
In the small-velocity limit, using the above equation one obtains the Newtonian value $-\frac{64G^5m^6\nu^2}{5b^4\pi p_\infty^4}$, in agreement with \eqref{eq:NewtonianSpectrum} below, and the 2.5PN correction $\frac{512G^5m^6\nu^2}{5b^4\pi}p_\infty$ due to radiation reaction, as already reported in \cite{DiVecchia:2022nna}.
As also remarked in that reference in the PN limit, the result \eqref{eq:ZFLextended} does not receive any correction due to interference between linear and nonlinear memory. One can see this for generic velocities by applying \eqref{eq:replace} to \eqref{eq:ZFL}, so that the cross-term involving nonlinear memory would read
\begin{equation}\label{eq:NLZFL}
\left(
\frac{dE}{d\omega}
\right)_\text{ZFL,NL}
=
    -\frac{4G}{\pi}
    \int_k \rho(k)
\sum_{a} p_a\cdot k \log\left(-\eta_a\,\frac{p_a\cdot k}{m_a\Lambda} \right),
\end{equation}
with $\Lambda$ an arbitrary energy scale introduced to regulate the collinear divergence in $D=4$. To leading order for small deflections,
\begin{equation}\label{eq:eventuallyzero}
\left(
\frac{dE}{d\omega}
\right)_\text{ZFL,NL}
\simeq
    -\frac{4G}{\pi}
    \int_k \rho(k)
Q\cdot k \log\frac{\tilde{u}_1\cdot k}{\tilde{u}_2\cdot k}\,,
\end{equation}
whose $\mathcal O(G^5)$ contribution vanishes since $\rho(k)$ is invariant under $b\cdot k\to-b\cdot k$ to leading order in $G$ owing to the reality of the tree-level amplitude (see \cite{Heissenberg:2024umh} for a similar mechanism at play in the angular momentum loss).
At order $\mathcal O(G^6)$, instead, the contribution \eqref{eq:NLZFL} due to nonlinear memory effect is expected to be nonzero and is not captured by  \eqref{eq:ZFL}, \eqref{eq:ZFLx}.

\subsection{\texorpdfstring{$\omega^2(\log\omega)^2$}{omega2 (log(omega)2)}}
The first subleading contribution in the frequency expansion arises at order $\omega^2(\log\omega)^2$ and is instead not Lorentz-invariant. It depends not only on the kinematic invariants $x_{ab}$ and on the masses, but also on the $z_a$ which we define by
\begin{equation}
\label{variablesz}
p_a^\mu = \eta_a \left( E_a, \vec{k}_a\right),\qquad
E_a=\frac{m_a}{2}\left(\frac{1}{z_a}+z_a\right),\qquad
\big|\vec{k}_a\big|=\frac{m_a}{2}\left(\frac{1}{z_a}-z_a\right),
\end{equation}
with $0<z_a\le1$. Focusing for simplicity on $2\to2$ kinematics for the background process, we find that it takes the following form
for generic masses and deflections
\begin{equation}\label{eq:spectrumNLO}
    \left(
\frac{dE}{d\omega}
\right)_{\omega^2 (\log\omega)^2}
=
    C
    +
    \sum_{i=1}^{4} B_i\,\log z_i
    +
    A_{13} \log x_{13}
    +
    A_{24} \log x_{24}
    +
    A_{23} \log x_{23}
    +
    A_{14} \log x_{14}\,.
\end{equation}
We provide the explicit expressions for the rational functions $C$, $B_{i}$ and $A_{ij}$ in terms of these variables in the ancillary file. 
Starting from the general expression \eqref{eq:spectrumNLO}, we can calculate the leading order in various limits and in different frames.

In the center-of-mass frame, 
\begin{equation}\label{eq:z1z2z3z4CM}
    z_1=z_4=\sqrt{\frac{x(m_2 x+m_1)}{m_1 x + m_2}}\,,
    \qquad
    z_2=z_3=\sqrt{\frac{x(m_1 x+m_2)}{m_2 x + m_1}}\,,
\end{equation}
for elastic $2\to2$ kinematics. 
    In the high-energy limit $s\sim Q^2\gg m^2_{1,2}$ (with $s$ the center-of-mass energy squared), using \eqref{eq:z1z2z3z4CM},  Eq.~\eqref{eq:spectrumNLO} reduces to 
\begin{equation}\label{eq:nlo_massless}
    \left(
\frac{dE}{d\omega}
\right)_{\omega^2 (\log\omega)^2}
=
s\,
\frac{16G^3}{\pi}
\left[ 
s
+
2 s \log\left(
1-\frac{Q^2}{s}
\right)
-2Q^2 \log\left(
\frac{s}{Q^2}-1
\right)
\right],
\end{equation}
where $p_1^\mu+p_4^\mu=Q^\mu=-p_2^\mu-p_3^\mu$,
in agreement with 
the results of \cite{Sahoo:2021ctw} for massless $2\to2$ scattering.
In the small-deflection limit, \eqref{eq:nlo_massless} yields \cite{Sahoo:2021ctw}
\begin{equation}\label{eq:nlo_massless_smallQ}
        \left(
\frac{dE}{d\omega}
\right)_{\omega^2 (\log\omega)^2}
=
\frac{16G^3}{\pi}\,
s^2
\left[ 
1
-
\frac{2Q^2}{s}
\log\frac{s}{Q^2}
+\cdots
\right].
\end{equation}
Expanding instead \eqref{eq:spectrumNLO} to leading $\mathcal O(G^3)$ for small deflections and generic velocities, we obtain
\begin{equation}
\begin{split}
    \left(
\frac{dE}{d\omega}
\right)^{\mathcal{O}(G^3)}_{\omega^2 (\log\omega)^2}
&=
\frac{16 G^3}{\pi}
\left(m_2 x+m_1\right) \left(m_1 x+m_2\right)\frac{\left(x^2+1\right)^2 \left(x^4-4 x^2+1\right)^2}{m_1 m_2 x^2 \left(x^2-1\right)^7}
   \\
   &\times
   \Big[m_1 m_2 \left(x^2-1\right) \left(m_1 m_2 x^2+2 \left(m_1^2+m_2^2\right) x+m_1
   m_2\right)
   \\
   &
   -2 m_2^3 x \left(m_1 x^2+m_2 x+m_1\right) \ell_1
   -2 m_1^3 x \left(m_2 x^2+m_1 x+m_2\right) \ell_2\Big]
   \end{split}
\end{equation}
with 
\begin{equation}
    x=x_{12}\,,
    \qquad
    \ell_1
    =
    \log \left(\frac{x \left(m_1 x+m_2\right)}{m_2
   x+m_1}\right),
   \qquad
   \ell_2
   =
   \log \left(\frac{x \left(m_2 x+m_1\right)}{m_1
   x+m_2}\right),
\end{equation}
or, in terms of $\sigma$,
\begin{subequations}
\begin{align}\label{eq:(omegalogomega)2G3}
    \left(
\frac{dE}{d\omega}
\right)^{\mathcal{O}(G^3)}_{\omega^2 (\log\omega)^2}
  &=
  s\,  \frac{4G^3}{\pi} \frac{\sigma^2(2\sigma^2-3)^2}{(\sigma^2-1)^3}\,\mathcal{H}(\sigma,m_1,m_2)\,,
  \\ 
  \label{eq:calH}
  \mathcal{H}(\sigma,m_1,m_2)
  &=\left[
2(s-m_1 m_2 \sigma)
+
\frac{m_2^2(2m_1\sigma+m_2)}{m_1\sqrt{\sigma^2-1}}\,\ell_1
+
\frac{m_1^2(2m_2\sigma+m_1)}{m_2\sqrt{\sigma^2-1}}\,\ell_2
  \right],
  \end{align}
\end{subequations}
where $s=m_1^2+2 m_1 m_2 \sigma+ m_2^2$. It is interesting to note that in the large-velocity limit, $\sigma\to\infty$, \eqref{eq:(omegalogomega)2G3} smoothly reduces to the massless result \eqref{eq:nlo_massless} expanded for small deflections as in \eqref{eq:nlo_massless_smallQ},
\begin{equation}
    \left(
\frac{dE}{d\omega}
\right)^{\mathcal{O}(G^3)}_{\omega^2 (\log\omega)^2}
\approx 
\frac{16 G^3}{\pi}\,s^2\,.
\end{equation}
This confirms the idea that when the massless result exists and is analytic at a given order in the deflection, then the corresponding PM result at that order in $G$ will smoothly reduce to it in the ultrarelativistic limit \cite{DiVecchia:2020ymx,DiVecchia:2021bdo}. 

Instead, in the frame where particle 1 is initially at rest, 
\begin{equation}\label{eq:z1z2z3z4REST}
    z_1=1\,,
    \qquad
    z_2=x\,,
    \qquad
    z_3=x_{13}\,,
    \qquad
    z_4=x_{14}
\end{equation}
for elastic $2\to2$ kinematics. Note that, to leading order in the probe limit $m_2^2 \ll m_1^2 \sim s$ for fixed $\sigma$, the two frames coincide, since the center-of-mass coincides with the position of the heavy particle. Correspondingly, \eqref{eq:z1z2z3z4CM} and \eqref{eq:z1z2z3z4REST} both reduce to $z_1=z_4=1$ and\footnote{This is because $Q_\text{1PM}\sim \mathcal O(m_1 m_2)$ and thus $\sigma_{14}\to 1$ as can be seen from \eqref{eq:sigma_explicit}.} $z_2=z_3=x$.
To leading order in $G$, in this frame we obtain
\begin{equation}\label{eq:spectrumNLOG3rest1}
\begin{split}
    \left(
\frac{dE}{d\omega}
\right)^{\mathcal{O}(G^3)}_{\omega^2 (\log\omega)^2}
&=
\frac{16 G^3 }{\pi}\, m_1^2 m_2^2 \\
&\times
\frac{(x^2+1)^2(x^4-4x^2+1)\left(x^6+9 x^4-9 x^2-1-12 \left(x^4+x^2\right)
   \log x\right)}{ 3x^2 \left(x^2-1\right)^7}
   \end{split}
\end{equation}
or, in terms of $\sigma$,
\begin{equation}\label{eq:nlo_rest}
    \left(
\frac{dE}{d\omega}
\right)^{\mathcal{O}(G^3)}_{\omega^2 (\log\omega)^2}
=
\frac{16 G^3}{\pi}\,m_1^2 m_2^2
\frac{\sigma ^2 \left(2 \sigma ^2-3\right)^2}{3 \left(\sigma ^2-1\right)^3}
\left(
\sigma^2+2-\frac{3\sigma\operatorname{arccosh}\sigma}{\sqrt{\sigma^2-1}}\right).
\end{equation}
One can check that, to leading order in the probe limit $m_2\ll m_1$, the results in the two frames \eqref{eq:(omegalogomega)2G3} and \eqref{eq:nlo_rest} precisely agree.

To leading order in the small-velocity limit, 
introducing
\begin{equation}
\sigma = \sqrt{1+p_\infty^2}\,,
\qquad
m = m_1 + m_2\,,
\qquad
\nu = \frac{m_1 m_2}{m^2}
\end{equation}
and letting $p_\infty\to0$,
the results for both observers \eqref{eq:(omegalogomega)2G3}, \eqref{eq:nlo_rest} at this order in $G$ reduce to
\begin{equation}
     \left(
\frac{dE}{d\omega}
\right)^{\mathcal{O}(G^3)}_{\omega^2 (\log\omega)^2}
=
\frac{32 G^3}{15 \pi p_\infty^2}\,m^4 \nu^2
+
\mathcal{O}(p_\infty^0)
\end{equation}
which matches Eq.~(2.11) of \cite{Kovacs:1978eu} (see also Eq.~\eqref{eq:NewtonianSpectrum} below).
The center-of-mass result \eqref{eq:(omegalogomega)2G3} has the following PN expansion up to subleading order,
\begin{equation}\label{eq:NLOspectrumPNG3}
     \left(
\frac{dE}{d\omega}
\right)^{\mathcal{O}(G^3)}_{\omega^2 (\log\omega)^2}
=
\frac{G^3m^4 \nu^2}{\pi p_\infty^2}\left[\frac{32}{15}+\left(-\frac{160}{21}+\frac{128}{105}\,\nu\right)p_\infty^2
+
\mathcal{O}(p_\infty^4)
\right],
\end{equation}
in perfect agreement with the result obtained from the frequency-domain analysis in \cite{Bini:2021jmj}. We also cross-checked this result by using the known multipole expansions of the waveform given in \cite{Georgoudis:2024pdz,Bini:2024rsy}.
Finally, \eqref{eq:nlo_rest} also agrees with the expression obtained very recently in \cite{Bini:2024ijq} up to the thirtieth order in the velocity. 

The result \eqref{eq:spectrumNLO} is an analytic function of $Q^2$ for small $Q^2$ and the next nontrivial PM order for the $\omega^2(\log\omega)^2$  portion of the spectrum is $\mathcal O(G^5)$. 
Its explicit expression reads as follows in the center-of-mass frame,
\begin{equation}\label{eq:(omegalogomega)2G5}
    \left(
\frac{dE}{d\omega}
\right)^{\mathcal{O}(G^5)}_{\omega^2 (\log\omega)^2}
=
- s \,\frac{2\sigma^2(2\sigma^2-3)^2}{(\sigma^2-1)^3}\frac{G^3 Q_\text{1PM}^2}{\pi} \, \mathcal{I}(\sigma) \,\,,
\end{equation}
where $s=m_1^2+2 m_1 m_2 \sigma + m_2^2$ and $\mathcal{I}(\sigma)$ is the function defined in \eqref{eq:mathcalI}.
It is rather remarkable that the only nontrivial dependence on the mass-ratio enters via $s$ and that the result is simply proportional to the leading PM order of the ZFL in Eq.~\eqref{eq:ZFL_PM}.
In particular, \eqref{eq:(omegalogomega)2G5} is significantly simpler than the previous order \eqref{eq:(omegalogomega)2G3}.

Considering the large-velocity limit 
$\sigma\to\infty$ of \eqref{eq:(omegalogomega)2G5} we obtain
\begin{equation}
    \left(
\frac{dE}{d\omega}
\right)^{\mathcal{O}(G^5)}_{\omega^2 (\log\omega)^2}
\approx
-
\frac{128 G^5}{b^2\pi}
s^3 
\log\frac{s}{m_1 m_2}
\end{equation}
so that the leading logarithm $\log s$ agrees with the corresponding one in \eqref{eq:nlo_massless_smallQ} after using $Q \approx \frac{2G s}{b}$ which holds in the massless case. One indeed expects divergences in the large-velocity limit when the massless result is non-analytic in the Newton constant, but logarithms should still match between the two (in principle different) regimes \cite{DiVecchia:2022owy,DiVecchia:2022nna}.

The result \eqref{eq:(omegalogomega)2G5} has the following PN expansion up to subleading order,
\begin{equation}\label{eq:NLOspectrumPNG5}
     \left(
\frac{dE}{d\omega}
\right)^{\mathcal{O}(G^5)}_{\omega^2 (\log\omega)^2}
=
-
\frac{G^5m^6 \nu^2}{\pi b^2 p_\infty^6}\left[\frac{64}{5}+\left(\frac{256}{35}+\frac{64}{5}\,\nu\right)p_\infty^2\right]\,.
\end{equation}
The first term here should agree with the NNLO small-$G$ limit of Eq.~(3.27) of \cite{Bini:2021jmj} for small frequencies. However, it is impractical to take the PM limit of that formula because of its dependence on $G$ via the indices of the Bessel functions. 
Nonetheless, one can easily obtain an independent  cross-check for it by expanding the time-domain waveform at Newtonian order. This leads to \eqref{eq:NewtonianSpectrum} below, which matches the first terms on the right-hand sides of both \eqref{eq:NLOspectrumPNG3} and \eqref{eq:NLOspectrumPNG5}.

Coming back to the rest frame of particle 1, expanding the general expression \eqref{eq:spectrumNLO} we find that the next correction to \eqref{eq:nlo_rest} takes the following form,
\begin{equation}\label{eq:nlo_restG5}
\begin{split}
&\left(
\frac{dE}{d\omega}
\right)^{\mathcal{O}(G^5)}_{\omega^2 (\log\omega)^2}
=
- s \,\frac{2\sigma^2(2\sigma^2-3)^2}{(\sigma^2-1)^3}\frac{G^3 Q_\text{1PM}^2}{\pi}\mathcal I(\sigma) \,
\\
&+
m_2\,
\frac{G^3 Q^2_\text{1PM}}{15\pi}\,\frac{2\sigma^2(2\sigma^2-3)^2}{(\sigma^2-1)^4}
\Bigg[
10 m_1 \sigma  \left(31-16 \sigma ^2\right)+m_2 \left(-18 \sigma ^4-29 \sigma ^2+32\right)
\\
&+\left(m_2
   \sigma  \left(4 \sigma ^2-3\right)+2 m_1 \left(4 \sigma ^4-6 \sigma ^2-3\right)\right) 
   \frac{15\operatorname{arccosh}\sigma}{\sqrt{\sigma^2-1}}
\Bigg]
\,\,.
   \end{split}
\end{equation}
Note that, in the probe limit $m_2\ll m_1$, the last two lines of \eqref{eq:nlo_restG5} become negligible (thanks to the extra power of $m_2$) and the result once again coincides with the one obtained in the center-of-mass frame \eqref{eq:(omegalogomega)2G5}.
The PN expansion of \eqref{eq:nlo_restG5} up to subleading order reads as follows,
\begin{equation}\label{eq:NLOspectrumPNG5rest}
     \left(
\frac{dE}{d\omega}
\right)^{\mathcal{O}(G^5)}_{\omega^2 (\log\omega)^2}
=
-
\frac{G^5m^6 \nu^2}{\pi b^2 p_\infty^6}\left[\frac{64}{5}
+
\frac{64}{7}\left(1-\frac{1}{5}\,\nu-\frac{1}{5}\sqrt{1-4\nu}\right)p_\infty^2\right]\,.
\end{equation}
The leading term here coincides with the one in \eqref{eq:NLOspectrumPNG5} and thus matches the correct Newtonian value in \eqref{eq:NewtonianSpectrum}.

As a further set of cross-checks, we also independently calculated the $\mathcal{O}(G^3)$ and $\mathcal{O}(G^5)$ expressions \eqref{eq:(omegalogomega)2G3}, \eqref{eq:nlo_rest}, \eqref{eq:(omegalogomega)2G5}, \eqref{eq:NLOspectrumPNG5rest} using the explicit PM-expanded soft waveforms collected in Subsection~\ref{PMwaveform}, finding agreement with the expressions derived from \eqref{eq:spectrumNLO} in this subsection.

Contributions due to the emission of gravitons and dissipative effects perturbing the $2\to2$ dynamics of the background process are expected to contribute to this observable starting from $\mathcal O(G^6)$, as is clear by applying \eqref{eq:replace} to \eqref{eq:NLOint} and recalling that $\rho\sim\mathcal{O}(G^3)$ to leading PM order. We leave their investigation for future work.

\subsection{Resummed spectrum in the center-of-mass frame}

Relying on the resummed soft waveform \eqref{eq:finalproposal2to2} for a $2\to2$ hard process, and plugging it into \eqref{eq:todointegrals}, we can obtain an expression for the energy spectrum $\frac{dE}{d\omega}$ that retains an exact dependence on the ``leading $\log$s'' (LL) and on the kinematics of the hard states. This calculation is particularly simple in the center-of-mass frame, where $E=\sqrt{s}$ is the total energy of the system and is angle-independent. In this way, the factors $\omega^{\pm i G E h(\sigma) \omega}$ are simply spectators in the integrals involved in \eqref{eq:todointegrals}, which we can evaluate  again as in Appendix~\ref{app:integrals}. Performing the calculation this frame, we thus find 
\begin{equation}
    \left(
\frac{dE}{d\omega}
    \right)_{\text{LL}}
    =
    F(\omega, m_1, m_2, \sigma; Q)\,,
\end{equation}
where the dependence on the deflection $Q$ enters via the invariants in \eqref{eq:sigma_explicit}.
We provide this general resummed expression $F(\omega,m_1,m_2,\sigma;Q)$ in the ancillary file.

It is then interesting to expand $F(\omega, m_1, m_2, \sigma; Q)$ for small deflections. In this way, we obtain
\begin{equation}\label{eq:resummedLeadingLogs}
\begin{split}
\left(
\frac{dE}{d\omega}
\right)_{\text{LL}}
&=
\left[\sin\left(G E h(\sigma) \omega\log\omega \right)\right]^2\,\frac{4G}{\pi}\,\mathcal{H}(m_1,m_2,\sigma)
\\
&+\cos\left(2G E h(\sigma) \omega\log\omega \right)\,\frac{GQ^2}{\pi}\,\mathcal{I}(\sigma) + \cdots 
\end{split}
\end{equation}
where $\mathcal{H}(\sigma,m_1,m_2)$ and $\mathcal{I}(\sigma)$ are the functions defined in \eqref{eq:calH} and \eqref{eq:mathcalI}. Higher-order terms in the $Q$-expansion \eqref{eq:resummedLeadingLogs} will be sensitive to nonlinear memory and radiative backreaction effects, which have been not taken into account in \eqref{eq:finalproposal2to2}. 

By rewriting the resummed formula \eqref{eq:resummedLeadingLogs} as  
\begin{equation}\label{eq:resummedLeadingLogsrew}
\begin{split}
\left(
\frac{dE}{d\omega}
\right)_{\text{LL}}
&=
\left[1-\cos\left(2G E h(\sigma) \omega\log\omega \right)\right]\,\frac{2G}{\pi}\,\mathcal{H}(m_1,m_2,\sigma)
\\
&+\cos\left(2G E h(\sigma) \omega\log\omega \right)\,\frac{GQ^2}{\pi}\,\mathcal{I}(\sigma) + \cdots 
\end{split}
\end{equation}
and expanding for small $G$, we see that the first line fixes all terms of order $G^{2n+1} (\omega\log\omega)^{2n}$ for $n=1,2,\ldots\,$, which are the leading PM contributions for all leading logs, 
\begin{equation}\label{eq:allorderLO}
\left(
\frac{dE}{d\omega}
\right)_{(\omega\log\omega)^{2n}}^{\mathcal{O}(G^{2n+1})}
=
\frac{(-1)^{n+1}}{(2n)!}\,\left(2G E h(\sigma) \omega\log\omega \right)^{2n}\,\frac{2G}{\pi}\,\mathcal{H}(m_1,m_2,\sigma)\,,
\end{equation}
thus reproducing the leading PM contribution of the $\omega^2(\log\omega)^2$ piece \eqref{eq:(omegalogomega)2G3} for $n=1$.  
Instead, the second line of \eqref{eq:resummedLeadingLogsrew} predicts all terms scaling as $G^{2n+3} (\omega\log\omega)^{2n}$ for $n=0,1,2,\ldots\,$, 
\begin{equation}\label{eq:allorderNLO}
\left(
\frac{dE}{d\omega}
\right)_{(\omega\log\omega)^{2n}}^{\mathcal{O}(G^{2n+3})}
=
\frac{(-1)^{n}}{(2n)!}\,\left(2G E h(\sigma) \omega\log\omega \right)^{2n}\,\frac{GQ^2_\text{1PM}}{\pi}\,\mathcal{I}(\sigma)\,,
\end{equation}
thus matching the leading PM order of the ZFL \eqref{eq:ZFL_PM} for $n=0$  and the subleading PM order of the $\omega^2(\log\omega)^2$ piece \eqref{eq:(omegalogomega)2G5} for $n=1$. 
In this way, the resummed formula generalizes the pattern exhibited by the ZFL and the $\omega^2(\log\omega)^2$ contributions, providing their all-order generalizations \eqref{eq:allorderLO}, \eqref{eq:allorderNLO}.

Vice-versa, in the high-energy limit, we find, recalling that $h(\sigma)\to2$ as $\sigma\to\infty$,
\begin{equation}
\begin{split}
\left(
\frac{dE}{d\omega}
\right)_{\text{LL}}
&= \frac{4G}{\pi}\,
[\sin(2G \sqrt{s}\, \omega\log\omega)]^2\,s
\\
&+
\frac{4G}{\pi}\,
\cos(4G \sqrt{s}\, \omega\log\omega)
\left[Q^2 \log \left(\frac{s}{Q^2}-1\right) - s\log\left(1-\frac{Q^2}{s}\right)\right].
\end{split}
\end{equation}
Expanding the trigonometric functions, this reproduces the ZFL \eqref{eq:ZFLUR} and the NLO \eqref{eq:nlo_massless} in the UR limit. 

\subsection{\texorpdfstring{$\omega^2\log\omega$}{omega2 log(omega)}}

In the previous sections, we discussed the leading and subleading contributions $\left(
\frac{dE}{d\omega}
\right)_{\text{ZFL}}$ and $\left(
\frac{dE}{d\omega}
\right)_{\omega^2(\log\omega)^2}$ to the low-frequency spectrum \eqref{eq:ZFexpandedSpectrum}, which are given by substituting the ``universal'' $\frac{1}{\omega}$, $\log\omega$ and $\omega(\log\omega)^2$ terms of the soft waveform \eqref{eq:wtildesoftresummed} into the integral \eqref{eq:todointegrals}.

Although they are not known non-perturbatively, we can use the explicit results of \cite{Georgoudis:2024pdz} for the tree-level $\omega^0$ and $\omega \log\omega$ contributions to $\tilde w^{\mu\nu}$ for the scattering of massive scalars in general relativity, substitute them into \eqref{eq:todointegrals} and evaluate the relevant angular integrals as in Appendix~\ref{app:integrals} to obtain the sub-subleading term $\left(
\frac{dE}{d\omega}
\right)_{\omega^2 \log\omega}$ to leading order $\mathcal{O}(G^3)$ in the PM expansion.

To write down the result in center-of-mass frame, we introduce the following notation for the mass-ratio,
\begin{equation}
q = \frac{m_2}{m_1}\,.
\end{equation}
Then we have
\begin{align}\label{eq:nunCM}
		&\left(
		\frac{dE}{d\omega}
		\right)_{\omega^2\log\omega}
		=
		\frac{4G^3 m_2^2 s \sigma (2\sigma^2-3))}{\pi q^2(\sigma^2-1)^3}
  \\
  \nonumber
        &\times
		\Bigg[
		2 P_0 +\frac{P_x \log x + P_{xq}\log\tfrac{x+q}{1+x q}}{q\sqrt{\sigma^2-1}} 
		-
		\frac{4}{q}(\sigma^2-1)\left(Q_x (\log x)^2+ Q_{xq}(\log x)\log\tfrac{x+q}{1+x q}\right)
	\\
 \nonumber
		&-\frac{1+q^2}{\sqrt{\sigma^2-1}} \,\left(4 \sigma ^4-6 \sigma ^2+1\right) 
  \left(
\operatorname{Li}_2\left(-\tfrac{q}{x}\right)-\operatorname{Li}_2\left(-qx\right)
		+(\log x)\log\frac{m_2^2}{q^2 s}
  \right)
		\Bigg]
\end{align}
where
\begin{subequations}
\begin{align}
P_0 &= -2 \left(q^2+1\right) \sigma ^3+q^2 \sigma -4 q \sigma ^4+4 q \sigma ^2+q+\sigma\,,\\
\begin{split}
P_x &=-6 \left(q^3+q\right)-8 q^2 \sigma ^5-10 q \left(q^2+1\right) \sigma ^4\\
&+15 q \left(q^2+1\right) \sigma ^2+\left(q^2-1\right)^2 \sigma -2 \left(q^4-6 q^2+1\right) \sigma ^3	\,,
\end{split}
\\
P_{xq} &= \left(1-q^2\right) \left(2 \left(q^2+1\right) \sigma ^3-\left(q^2+1\right) \sigma +2 q \sigma ^4+q \sigma ^2-2 q\right),\\
Q_x &= q^4+q^3 \sigma +q \sigma +1\,,\\
Q_{xq} &= \left(q^2-1\right) \left(q^2+q \sigma +1\right).
\end{align}
\end{subequations}
In the PN limit,
\begin{equation}\label{eq:nunPN}
	\left(
	\frac{dE}{d\omega}
	\right)_{\omega^2\log\omega}
	=
	\frac{G^3 m^4 \nu ^2}{\pi p_\infty^2}	
 \left[
    -\frac{64}{5}
    +\left(\frac{512}{35}
	-\frac{512 \nu }{105}\right)p_\infty^2
	+\mathcal O(p_\infty^4)
	\right]
\end{equation}
in perfect agreement with the expression obtained from the PN-expanded waveforms calculated in \cite{Georgoudis:2024pdz}. Let us note that a nontrivial dependence on the mass-ratio is indeed a common feature of observables involving the frequency defined in the center of mass \cite{Dlapa:2024cje}.

In the rest frame of particle 1, we find instead
\begin{equation}\label{eq:nunrest1}	\begin{split}
&\left(
\frac{dE}{d\omega}
\right)_{\omega^2\log\omega}
=
-\frac{4G^3m_1^2m_2^2\sigma^2}{3\pi(\sigma^2-1)^3}
\Bigg[
(2\sigma^2-3)
\Bigg(
24(\sigma^2-1)
(\operatorname{arccosh}\sigma)^2
\\
&
-
(16 \sigma^6+12\sigma^4-42\sigma^2+17)
\frac{\operatorname{arccosh}\sigma}{\sigma\sqrt{\sigma^2-1}}
\Bigg)
+
16 \sigma^6-70\sigma^2+51
\Bigg].
	\end{split}
\end{equation}
One can check that indeed \eqref{eq:nunCM} reduces to \eqref{eq:nunrest1} to leading order in the probe limit, $q\to0$, and that the leading PN limit of \eqref{eq:nunrest1} agrees with the first term in \eqref{eq:nunPN} as it should.
Finally, \eqref{eq:nunrest1} also agrees with the result of \cite{Bini:2024ijq}, which includes up to the thirtieth order in the velocity.

\section{Time domain}
\label{sec:timedomain}

As reviewed in Appendix~\ref{app:FTtomega}, the small-frequency expansion \eqref{eq:wtildesoft} corresponds to the following large-time one via Fourier transform,
\begin{equation}\label{eq:Apmdeftw}
    w^{\mu\nu} \sim 
    {A}_0^{(\pm)\mu\nu} 
    +\sum_{n=1}^\infty \frac{{A}_{n}^{(\pm)\mu\nu}}{t^n}\,(\log |t|)^{n-1}
    + \cdots
    \qquad
    \text{as }t\to\pm\infty
\end{equation}
where $t$ denotes retarded time and\footnote{For the time being, we follow the splitting between $t\to+\infty$ and $t\to-\infty$ contributions which is more natural for the formulas in Subsection~\ref{ssec:rewritten}. Different choices of $i0$ prescriptions in frequency domain allow one to move terms from $A_n^{(-)\mu\nu}$ to $A_n^{(+)\mu\nu}$\cite{Sahoo:2020ryf,Sahoo:2021ctw} while leaving $A_n^{\mu\nu}$ invariant (see Appendix~\ref{app:FTtomega} and Section~\ref{ssec:resummedtime}).}
\begin{equation}
    A_n^{\mu\nu} = {A}_n^{(+)\mu\nu} - {A}_n^{(-)\mu\nu}\,.
\end{equation}

\subsection{Newtonian limit}
\label{NewtonianLimit}

In this section, we discuss the soft-expanded waveform to leading order in the post-Newtonian limit, in which $p_\infty\to0$ for fixed $\frac{G m}{p_\infty^2 b}$. For simplicity, we will work with the convention $G=m=1$. One can reinstate these constants, if needed, by noting that powers of $p_\infty^{-2}$ must be accompanied by powers of $G$ at a given PN order and by using dimensional analysis.

Choosing Cartesian coordinates $x^1$, $x^2$ on the scattering plane, the Keplerian trajectory reads (see~\cite{Cho:2018upo,Bini:2020hmy,Bini:2021jmj} and references therein)
\begin{subequations}\label{eq:Kepleriantrajectory}
    \begin{align}
    \label{eq:relationsolvev}
x^0 &= \bar{a}_r^{\frac{3}{2}} (e_r\sinh v - v)\,,\\
x^1 &= \bar{a}_r \left( e_r - \cosh v \right),\\
x^2 &= \bar{a}_r \sqrt{e_r^2-1} \sinh v\,,\\
x^3 &= 0\,,
    \end{align}
\end{subequations}
where $v$ is a parameter and $e_r$ is the eccentricity. 
To Newtonian accuracy, the parameters $\bar{a}_r$, $e_r$ are linked to velocity variable $p_\infty = \sqrt{\sigma^2-1}$ and to the initial impact parameter $b$ by
\begin{equation}\label{eq:bararer}
    \bar{a}_r = \frac{1}{p_\infty^2}\,,
    \qquad
    e_r =\sqrt{1+(bp_\infty^2)^2} = b p_\infty^2 + \frac{1}{2b p_\infty^2}+\mathcal{O}\left(\tfrac{1}{b^3p_\infty^6}\right)\,.
\end{equation}
The trajectory \eqref{eq:Kepleriantrajectory} determines the waveform, to leading order in the post-Newtonian approximation, via the quadrupole formula, (with Latin indices $i,j,k,\ell,\ldots$ taking values $1,2,3$)
\begin{equation}\label{eq:quadrupole}
    w_{ij}^{\text{TT}} = \frac{1}{2!}\, \Pi_{ijk\ell}\,U_{k\ell} \,,
    \qquad
    U_{ij} = \nu \left( x_{\langle i} x_{j\rangle}\right)^{(2)},
\end{equation}
where, $\Pi_{ijk\ell}$ is the purely spatial transverse-traceless (TT) projector, angular brackets denote the symmetric trace-free projection and the superscript $(2)$ stands for the second derivative with respect to $x^0$. 

Note that the trajectory \eqref{eq:Kepleriantrajectory} is more naturally expressed in terms of the parameter $v$, rather than the coordinate time $x^0$.
However, the relation between $x^0$ and $v$ can be solved for $v$ perturbatively for large $|x^0|$. Indeed, in the limit $x^0\to\pm \infty$ and $v\to\pm\infty$, Eq.~\eqref{eq:relationsolvev} can be recast as follows
\begin{equation}\label{eq:wysolvev}
x \simeq |y| + \frac{2}{e_r}\,\log x\,,
\end{equation}
with
\begin{equation}\label{eq:xasy}
    x = e^{|v|}\,,\qquad y = \frac{2 x^0}{e_r \bar a_r^{\frac{3}{2}}}\,.
\end{equation}
The solution of \eqref{eq:wysolvev} can be written down formally in terms of nested logarithms by iterating the right-hand side, $x=|y|+\frac{2}{e_r}\,\log\left(|y|+\frac{2}{e_r}\,\log(|y|+\cdots)\right)$. However, since we are focusing here on the leading logarithmic dependence on time, in order to capture the terms displayed in \eqref{eq:Apmdeftw}, we can safely stop at the first iteration,
\begin{equation}\label{eq:solvedv}
    x \simeq |y| + \frac{2}{e_r}\,\log\left(|y|+\frac{2}{e_r}\,\log|y|\right). 
\end{equation}

Contracting $w_{\mu\nu}$ with a reference vector $\varepsilon^{\mu}=(0,\varepsilon_1,\varepsilon_2,\varepsilon_3)$ such that $\varepsilon\cdot n=0$, $\varepsilon\cdot \varepsilon = 0$,
substituting \eqref{eq:Kepleriantrajectory} into \eqref{eq:quadrupole},
and using \eqref{eq:solvedv} (see Appendix~\ref{app:newt} for more details), one obtains the following Newtonian-level expression for the (retarded) time-domain waveform at large $|t|$,
\begin{equation}\label{eq:ewededuced}
\varepsilon\cdot w\cdot\varepsilon
\simeq
\frac{\nu}{\bar{a}_r}\, \left[ \frac{\varepsilon_1^2}{e_r^2} 
\mp \frac{2 \varepsilon_1 \varepsilon_2}{e_r}\sqrt{1-\frac{1}{e_r^2}} 
+
\varepsilon_2^2 \left(1-\frac{1}{e_r^2}\right)\right]
\left(1+\frac{1}{\tau}\right),
\qquad
\text{as }t\to\pm\infty\,,
\end{equation}
where we have introduced the variable 
\begin{equation}
    \label{eq:tau}
    \tau = \frac{|t|}{\bar{a}_r^{\frac{3}{2}}}+\log|t|\,.
\end{equation}
Although the $\frac{1}{\tau}$ behavior in \eqref{eq:ewededuced} is sometimes referred to as ``tail'', it should be kept distinct from tail effects which in the PN literature appear as integrals over the past history of the binary. The latter effects only appear at 1.5PN order (see e.g.~Section 2.4.2 of \cite{Blanchet:2013haa}) and are associated to rescattering phenomena, encoded in $\mathcal{W}_\text{phase}$ in our approach, while the former are related to $\mathcal{W}_{\text{reg}}$ and are already visible at 0PN.

From the Taylor series of the last factor in \eqref{eq:ewededuced}, where $\tau$ is given by the simple expression \eqref{eq:tau}, it is straighforward to deduce the coefficients in \eqref{eq:Apmdeftw}. 
For instance, letting $A_n^{(\pm)} =\varepsilon_\mu A_n^{(\pm)\mu\nu}\varepsilon_\nu$, we find the coefficients associated to $\frac{1}{\omega}$, $\log\omega$ and $\omega(\log\omega)^2$,
\begin{subequations}\label{eq:ApmNewt}
\begin{align}
    A_0^{(\pm)} 
    &=
    \frac{\nu  \varepsilon _1^2}{e_r^2 \bar{a}_r}
    \mp\frac{2 \nu  \varepsilon _2 \varepsilon _1 \sqrt{e_r^2-1}}{e_r^2 \bar{a}_r}
    +\frac{\nu  \varepsilon _2^2 \left(e_r^2-1\right)}{e_r^2 \bar{a}_r}\,,
    \\
    A_1^{(\pm)} 
    &=
    \pm
    \frac{\nu  \varepsilon _1^2 \sqrt{\bar{a}_r}}{e_r^2}
    -\frac{2 \nu  \varepsilon _2 \varepsilon _1 \sqrt{\left(e_r^2-1\right) \bar{a}_r}}{e_r^2}
    \pm\frac{\nu  \varepsilon _2^2 \left(e_r^2-1\right) \sqrt{\bar{a}_r}}{e_r^2}\,,
    \\
    A_2^{(\pm)} 
    &=
    -\frac{\nu  \varepsilon _1^2 \bar{a}_r^2}{e_r^2}
    \pm\frac{2 \nu  \varepsilon _1 \varepsilon _2 \sqrt{e_r^2-1} \bar{a}_r^2}{e_r^2}
    -\frac{\nu  \varepsilon _2^2 \left(e_r^2-1\right) \bar{a}_r^2}{e_r^2}
\end{align}
\end{subequations}
so that
\begin{subequations}\label{eq:ANewt}
\begin{align}
\label{eq:A00PN}
A_0 
    &=
    -\frac{4 \nu  \varepsilon _1 \varepsilon _2 \sqrt{e_r^2-1}}{e_r^2 \bar{a}_r}\,,
    \\
\label{eq:A10PN}
A_1 
    &=
    \frac{2 \nu  \varepsilon _1^2 \sqrt{\bar{a}_r}}{e_r^2}+\frac{2 \nu  \varepsilon _2^2 \left(e_r^2-1\right) \sqrt{\bar{a}_r}}{e_r^2}\,,
\\
\label{eq:A20PN}
A_2 
    &=
    \frac{4 \nu  \varepsilon _1 \varepsilon _2 \sqrt{e_r^2-1} \bar{a}_r^2}{e_r^2}\,.
\end{align}
\end{subequations}
One can check that \eqref{eq:ANewt} precisely agree with the corresponding general expressions \eqref{eq:A0A1A2} expanded to Newtonian order for small $p_\infty$, using \eqref{A} with $Q_1^\mu \simeq Q^\mu \simeq -Q_2^\mu$
and the Newtonian impulse given by
\begin{equation}\label{eq:NewtonianImpulseandAngle}
    \Theta = 2 \arctan\frac{1}{\sqrt{e_r^2-1}}\,,
    \qquad
    Q = \frac{2\nu}{\sqrt{\bar{a}_r}} \sin\frac{\Theta}{2} = \frac{2\nu}{e_r\sqrt{\bar{a}_r}}\,,
\end{equation}
noting in particular that 
\begin{equation}
b_e^i = b_e\,(1,0,0)\,,
\qquad
\tilde{p}_1^{\,i} =  \tilde{p} \, (0,1,0)=-\tilde{p}^{\,i}_2
\end{equation}
in the present conventions, and at Newtonian level
\begin{equation}
\tilde{p}=\frac{\tilde{m}_1\tilde{m}_2\sqrt{\tilde{\sigma}^2-1}}{\sqrt{s}}
=\frac{\nu}{e_r \sqrt{\bar{a}_r}}\,\sqrt{e_r^2-1}\,.
\end{equation}
Similarly, one can check that the further small-$G$ expansion of the above Newtonian formulas agrees with the $p_\infty\to0$ limit of the expressions for the PM-expanded soft factors given in Subsections~\ref{PMwaveform}, as well as with the results in \cite{Bini:2024ijq}.

From \eqref{eq:ewededuced}, it is easy to include $A_{\ell}$ for $\ell\ge3$ as well. For instance, the next leading logarithm has the following coefficients,
\begin{equation}
    A_3^{(\pm)} 
    =
    \pm
    \frac{\nu  \varepsilon_1^2 \bar{a}_r^{7/2}}{e_r^2}
    -\frac{2 \nu  \varepsilon_1 \varepsilon_2 \sqrt{e_r^2-1} \bar{a}_r^{7/2}}{e_r^2}
    \pm\frac{\nu  \varepsilon_2^2 \left(e_r^2-1\right) \bar{a}_r^{7/2}}{e_r^2}\,,
\end{equation}
and thus
\begin{equation}
\label{eq:A30PN}
    A_3 
    =
    \frac{2 \nu  \varepsilon_1^2 \bar{a}_r^{7/2}}{e_r^2}+\frac{2 \nu  \varepsilon_2^2 \left(e_r^2-1\right) \bar{a}_r^{7/2}}{e_r^2}
    \,,
\end{equation}
which agrees with the Newtonian limit of \eqref{eq:A3conjecture}.
Finally, from the all-order expansion of the factor $\frac{1}{\tau}$ in \eqref{eq:ewededuced}, we find the following general results,
\begin{equation}\label{eq:A2kA2kp1Conjecture}
\begin{cases}
    A_{2k} 
    =
    \dfrac{4\nu \varepsilon_1\varepsilon_2\sqrt{e^2_r-1}}{e_r^2}\,\bar{a}_r^{-1+3k}
    & \text{for }k\ge1
    \\[10pt]
    A_{2k+1} 
    =
    2\nu
    \left(
    \dfrac{\varepsilon_1^2}{e_r^2}
    +\dfrac{\varepsilon_2^2 \left(e_r^2-1\right)}{e_r^2}
    \right)\bar{a}_r^{\frac{1}{2}+3k}
    \qquad
    &\text{for }k\ge0\,,
\end{cases}
\end{equation}
while $A_0$ is given in \eqref{eq:A00PN}.
We can also check that the sequences in \eqref{eq:A2kA2kp1Conjecture} are in perfect agreement with the Newtonian limit of the general expression \eqref{eq:proposalANYk}, by noting that the latter reduces to the following one for $\ell \ge 1$,
\begin{equation}\label{eq:alternating}
    \varepsilon \cdot a_\ell \cdot \varepsilon 
    =
    \bar{a}_r^{\frac{3\ell}{2}}\left[ 
    (-1)^\ell
    \varepsilon\cdot B(p_i)\cdot \varepsilon-\varepsilon\cdot B(p_f)\cdot \varepsilon
    \right]
\end{equation}
with
\begin{equation}\label{eq:exprBB}
    \varepsilon\cdot B(p_i)\cdot \varepsilon
    =
    \frac{\nu}{\bar{a}_r}
    \left(
    \frac{\varepsilon_1}{e_r}
    +
    \frac{\varepsilon_2}{e_r}\, \sqrt{e_r^2-1}
    \right)^2,
    \qquad
    \varepsilon\cdot B(p_f)\cdot \varepsilon
    =
    \frac{\nu}{\bar{a}_r}
    \left(
    \frac{\varepsilon_1}{e_r}
    -
    \frac{\varepsilon_2}{e_r}\, \sqrt{e_r^2-1}
    \right)^2\,.
\end{equation}
The alternating sign in \eqref{eq:alternating} is responsible for the emergence of the two structures in \eqref{eq:A2kA2kp1Conjecture} when $\ell=2k$ or $\ell=2k+1$
(note that $a_\ell^{\mu\nu}$ provides the dominant contribution to $A_\ell^{\mu\nu}$ in the Newtonian limit).

One can then plug the quadrupole $U_{ij}$ given by \eqref{eq:ANewt} into the Einstein quadrupole formula, 
\begin{equation}
 \left(
\frac{dE}{d\omega}
\right)^{\text{Newtonian}}
    =
    \frac{1}{5\pi}\,\omega^2\mathrm{U}^\ast_{ij} \mathrm{U}_{ij}\,,
    \qquad
    \mathrm{U}_{ij}
    =
    \int_{-\infty}^{+\infty} e^{i\omega t} U_{ij}\, dt\,,
\end{equation}
to obtain the leading estimate of the energy spectrum for small velocity.
By doing so, one finds 
\begin{equation}\label{eq:NewtonianSpectrumExact}
   \begin{split}
 \left(
\frac{dE}{d\omega}
\right)^{\text{Newtonian}}
&=
\frac{32\nu^2}{5 \bar{a}_r^2\pi e_r^4} (e_r^2-1)
\\
&+
\frac{32 \bar{a}_r \nu^2}{15 \pi e_r^4} \left[e_r^4-6(e_r^2-1)\right]
\omega^2 (\log\omega)^2
+
\mathcal O(\omega^2 \log\omega)\,.
   \end{split}
\end{equation}
Therefore, expanding \eqref{eq:NewtonianSpectrumExact} for large $b$ by means of \eqref{eq:bararer},
\begin{equation}\label{eq:NewtonianSpectrum}
   \begin{split}
 \left(
\frac{dE}{d\omega}
\right)^{\text{Newtonian}}
&=
\frac{\nu^2}{\pi b^2} 
\left(
\frac{32}{5} - \frac{64}{5 p_\infty^4 b^2}
+
\mathcal O \left(\tfrac{1}{p_\infty^8 b^4}\right)
\right)
\\
&+
\frac{\nu^2}{\pi p_\infty^2} 
\left(
\frac{32}{15} - \frac{64}{5 p_\infty^4 b^2}
+
\mathcal O \left(\tfrac{1}{p_\infty^8 b^4}\right)
\right)
\omega^2 (\log\omega)^2
+
\mathcal O(\omega^2 \log\omega)\,.
   \end{split}
\end{equation}

\subsection{Resummed \texorpdfstring{$2\to2$}{2->2} waveform for generic velocities}
\label{ssec:resummedtime}

Employing \eqref{eq:resum2omegaPM}, \eqref{eq:resum2tPM}, we can find the Fourier transform to time-domain\footnote{We are particularly grateful to Ali Seraj for very stimulating discussions on this topic.} of the waveform  \eqref{eq:finalproposal2to2} capturing the leading logarithms.
To this end, we need to assign suitable $\pm i0$ prescriptions, which effectively amounts to distributing the terms between $t\to\pm\infty$.
Following \cite{Sahoo:2018lxl,Saha:2019tub,Sahoo:2020ryf,Sahoo:2021ctw,Sen:2024bax}, we consider
\begin{equation}
\label{eq:finalproposal2to2PPPM}
\tilde{w}
=
\frac{i}{E \omega_+}
\, \omega_+^{2iGE\omega}
\left[ 
\omega_+^{-iGE \omega h(\sigma)}
B(p_f)
-
\omega_-^{iGE \omega h(\sigma)}
B(p_i)
\right]
+\cdots
\end{equation}
where we introduced $\omega_\pm= \omega\pm i0$,
we focused for simplicity on the center-of-mass frame and we saturated all indices with the reference vector $\varepsilon^\mu$ defined above \eqref{eq:ewededuced},
\begin{equation}
\tilde{w}
=\varepsilon\cdot\tilde{w}\cdot\varepsilon\,,\qquad
B(p_f)
=\varepsilon\cdot B(p_f)\cdot\varepsilon\,,\qquad
B(p_i)
=\varepsilon\cdot B(p_i)\cdot\varepsilon\,.
\end{equation}
In this way, applying \eqref{eq:resum2omegaPM}, \eqref{eq:resum2tPM}, we find the following result for the time-domain waveform, $w=\varepsilon\cdot w\cdot \varepsilon$, at leading log accuracy
\begin{subequations}\label{eq:wsoftresummedNLtime}
\begin{align}
\label{eq:wsoftresummedNLtimeplusnew}
\begin{split}
    {w}
    &\sim
    \frac{1}{E}\left(
B(p_f) - B(p_i) 
\right)
\\
&+
    \frac{G(2-h(\sigma))\,B(p_f)}{t+GE(2-h(\sigma))\log t} 
    -
    \frac{2G\,B(p_i)}{t+GE(2+h(\sigma))\log t} 
    +\cdots
    \,,
    \qquad
    \text{as }t\to+\infty
\end{split}
    \\
\label{eq:wsoftresummedNLtimeminusnew}
    {w}
    &\sim
    \,\frac{G h(\sigma)\,B(p_i)}{t+GE(2+h(\sigma))\log|t|} 
    +\cdots
    \,,
    \qquad
    \text{as }t\to-\infty
\end{align}
\end{subequations}
Note that 
\begin{equation}
\frac{1}{E}\left(
B^{\mu\nu}(p_f) - B^{\mu\nu}(p_i) 
\right)
=
- \left(
\frac{p_1^\mu p_1^\nu}{p_1\cdot n}
+
\frac{p_2^\mu p_2^\nu}{p_2\cdot n}
+
\frac{p_3^\mu p_3^\nu}{p_3\cdot n}
+
\frac{p_4^\mu p_4^\nu}{p_4\cdot n}
\right)
\end{equation}
consistently with the memory effect, although this contribution can be redistributed between $t\to\pm\infty$ by performing a BMS supertranslation (see e.g.~\cite{Strominger:2014pwa,Veneziano:2022zwh}). 
Noting that, as $p_\infty\to0$, the contributions involving $h(\sigma)$ dominate thanks to 
\begin{equation}
h(\sigma) \sim - \frac{1}{p_\infty^3} \sim - \bar{a}_r^{3/2}\,,
\end{equation}
Eqs.~\eqref{eq:wsoftresummedNLtime} reduce to \eqref{eq:ewededuced} at Newtonian level in view of \eqref{eq:exprBB} (up to the supertranslation in \cite{Veneziano:2022zwh}). 
The leading PM order for the $1/t$ tail predicted by \eqref{eq:wsoftresummedNLtime},
\begin{subequations}\label{}
\begin{align}
\label{}
    {w}
-
    \frac{1}{E}\left(
B(p_f) - B(p_i) 
\right)
&\sim
-\frac{Gh(\sigma)}{t}\, B(p_i)
+\mathcal{O}(G^2)
    \,,
    \qquad
    \text{as }t\to+\infty
    \\
\label{}
    {w}
    &\sim
    +\frac{G h(\sigma)}{t}\,B(p_i) 
    +\mathcal{O}(G^2)
    \,,
    \qquad
    \text{as }t\to-\infty
\end{align}
\end{subequations}
matches \cite{Sahoo:2018lxl,Saha:2019tub,Sahoo:2020ryf,Sahoo:2021ctw} and  also agrees (for generic $\sigma$) with the explicit results for the leading PM waveform in time domain \cite{Kovacs:1978eu,Jakobsen:2021lvp,Jakobsen:2021smu}.
It will be interesting to perform further checks at higher PM orders against the explicit prediction provided by \eqref{eq:wsoftresummedNLtime} for the terms behaving as $\frac{G^n}{t^n}\,(\log|t|)^{n-1}$.

\section{Conclusions}
\label{sec:concl}

In this work, we used the classical soft graviton theorems of Refs.~\cite{Sahoo:2018lxl,Laddha:2019yaj,Saha:2019tub,Sahoo:2020ryf,Sahoo:2021ctw,Ghosh:2021bam} to obtain the energy spectrum characterizing gravitational wave emissions sourced by encounters of compact objects, up to NLO, i.e.~order $\omega^2(\log\omega)^2$ in the frequency. This allowed us to study its behavior both in the ultrarelativistic limit and in the PM regime, providing explicit expressions up to $\mathcal{O}(G^5)$ in the PM expansion, while at order $\mathcal{O}(G^6)$ nonlinear  effects due to soft emissions by hard gravitons start contributing. At order $\mathcal{O}(G^3)$, we also computed the $\omega^2\log\omega$ contribution to the spectrum by using the explicit tree-level expression for the $\omega^0$ and $\omega\log\omega$ waveforms at tree level \cite{Georgoudis:2023eke}.

It was observed in \cite{Sahoo:2021ctw} that, while the nonlinear memory effect appearing at $\frac{1}{\omega}$ involves a nontrivial integral over the phase space density of emitted gravitons, at orders $\log\omega$ and $\omega(\log\omega)^2$ nonlinear effects can be expressed in terms of the mismatch between the initial and final momenta of the massive states. The cancellations leading to this simplification are reminiscent of those ensuring the absence of collinear singularities in the infrared divergences of gravitational amplitudes \cite{Weinberg:1965nx}. Indeed, classical soft theorems can be also deduced from the infrared divergences of amplitudes involving gravitons by applying Eq.~\eqref{resummedWeinb} \cite{Krishna:2023fxg,Alessio:2024wmz}. It will be interesting to further explore the interplay between these ingredients in the future, in order to better understand the link between the amplitude and the waveform \cite{Kosower:2018adc,Cristofoli:2021vyo}.

Based on the structures emerging in the first few contributions, $\frac{1}{\omega}$, $\log\omega$ and $\omega(\log\omega)^2$ to the soft expansion of the waveform, we also provided a conjecture for the expression of the leading logarithmic contributions $\omega^{n-1}\,(\log\omega)^n$ for any $n\ge3$ for generic kinematics of the hard states. While in this work we only provided a derivation of these expressions for generic $n$ in the Newtonian limit for a $2\to2$ background process, it will be important to perform further checks and revisit this proposal in the spirit of \cite{Sahoo:2018lxl,Laddha:2019yaj,Saha:2019tub,Sahoo:2020ryf,Sahoo:2021ctw,Ghosh:2021bam} for more general kinematics and multiplicities. One could start by testing \eqref{eq:finalproposalNtoM} for a system of $N$ particles in the PN approximation \cite{Kidder:2007rt}, and for instance in the case of $3\to3$ scattering we find that $a_3^{\mu\nu}$ is already nontrivial at Newtonian order $G^3/p_\infty^7$, as it happened for the $2\to2$ case. The prospect of gaining better control of such all-order leading logarithms is particularly interesting in connection with their universality and their link with asymptotic symmetries \cite{Agrawal:2023zea,Choi:2024ygx}.

\subsection*{Acknowledgments}
We would like to thank Shreyansh Agrawal, Laura Donnay, Alessandro Georgoudis, Vasco Gon\c{c}alves, Julio Parra-Martinez, Andrea Puhm, Rodolfo Russo, Biswajit Sahoo, Ali Seraj, Piotr Tourkine, Gabriele Veneziano for helpful discussions.
The research of F.~A. (P.~D.~V.) is fully (partially) supported by the Knut and Alice Wallenberg Foundation under grant KAW 2018.0116.
C.~H. is supported by UK Research and
Innovation (UKRI) under the UK government’s Horizon Europe funding guarantee [grant EP/X037312/1 ``EikoGrav: Eikonal Exponentiation and Gravitational
Waves''].  

\appendix

\section{Useful integrals}
\label{app:integrals}

\subsection{Generic kinematics}

In Section~\ref{enspect}, the various contributions to the soft spectrum of emitted energy in \eqref{eq:ZFexpandedSpectrum} are given in terms of integrals over the angles $(\theta,\phi)$ appearing in \eqref{eq:LOint} and \eqref{eq:NLOint}. They can be reduced to integrals of the form:
\begin{align}
\mathcal{I}^{i_i\cdots i_n}_{m}=&\int\frac{d\Omega}{4\pi}\frac{n^{i_i}\cdots n^{i_n}}{\prod_{k=1}^m(p_{a_k}\cdot n)}\,,
\end{align}
with $m\leq 2$ and $n\leq 2$.
As instructive examples, let us analyse here explicitly the scalar cases with $n=0$ and $m=1,2$. Using rotational invariance, we  parametrize the momenta as $k_a^{\mu}=(E_a,0,0,k_a)$ and $k_b^{\mu}=(E_b,0,k_b^{y},k_b^{z})$. However, we find it useful to trade these variables
for $\sigma_{ab}$ and $z_{a}$, $z_b$ defined in \eqref{eq:sigmadef} and \eqref{variablesz}.
Let us start with $m=1$
\begin{align}
\label{ex1}
\mathcal{I}_1=\int\frac{d\Omega}{4\pi}\frac{1}{p_{a}\cdot n}=\eta_{a}\int\frac{d\Omega}{4\pi}\frac{1}{k_{a}\cdot n}\,.
\end{align}
Letting $x=\cos\theta$, we get 
\begin{equation}
\mathcal{I}_1 
=
-
\frac{\eta_a}{m_a}
\int_{-1}^{1}
\frac{z_a\, dx}{1+z_a^2-(1-z_a^2)x}
=
\frac{2\eta_a}{m_a}
\frac{z_a\log z_a}{1-z_a^2}
\end{equation}
Notice how this result explicitly depends on $z_a$ and hence on the particular frame chosen. For $m=2$ we have
\begin{align}
\label{ex2}
\mathcal{I}_2=\int\frac{d\Omega}{4\pi}\frac{1}{(p_{a}\cdot n)(p_{b}\cdot n)}=\eta_{a}\eta_{b}\int\frac{d\Omega}{4\pi}\frac{1}{(k_{a}\cdot n)(k_{b}\cdot n)}\,,
\end{align}
in terms of which the result of the integral in \eqref{ex2} can easily be recast as
\begin{align}
\mathcal{I}_2=\frac{\eta_{a}\eta_{b}}{4m_{a}m_{b}}\frac{\log(\frac{1-2\sigma_{ab}^2-2\sigma_{ab}\sqrt{\sigma_{ab}^2-1}}{1-2\sigma_{ab}^2+2\sigma_{ab}\sqrt{\sigma_{ab}^2-1}})}{\sqrt{\sigma_{ab}^2-1}}=\frac{\eta_{a}\eta_{b}}{m_{a}m_{b}}\frac{\mathrm{arccosh}\sigma_{ab}}{\sqrt{\sigma_{ab}^2-1}}\,.
\end{align}
Notice that, differently from \eqref{ex1}, the result of this integral (appearing in the ZFL) is actually Lorentz invariant and depends only on the kinematic invariants $\sigma_{ab}$ and on the masses, but not on $z_{a}$. The results for the other vector and tensor integrals with $n\neq 0$ can be obtained using Passarino--Veltman reduction in terms of the scalar ones discussed here.

\subsection{Small deflections}

After performing the PM expansion of the soft waveforms as in Subsection~\ref{PMwaveform}, all the integrals appearing in the energy spectrum in Section~\ref{enspect} reduce to the  master integrals
\begin{align}
\mathcal{I}^{(j)}(\alpha,\beta)=\int\frac{d\Omega}{4\pi}\frac{(b\cdot n)^j}{(p_1\cdot n)^{\alpha}(p_2\cdot n)^{\beta}},
\end{align}
with $j=0,1,2$.

\subsubsection*{Frame attached to particle 1}

In this frame, particle 1 is initially at rest and therefore the two incoming momenta are $p_1^{\mu}=-m_1 v_1^{\mu}$ and $p_2^{\mu}=-m_2 v_2^{\mu}$ with 
\begin{align}
\label{par1}
v_1^{\mu}=(1,0,0,0),\qquad v_2^{\mu}=(\sigma,0,0,\sqrt{\sigma^2-1}).
\end{align}
Defining $\sigma_{\pm}=\sigma\pm\sqrt{\sigma^2-1}$, we have
\begin{align}
\mathcal{I}^{(0)}(\alpha,\beta)
&=\frac{(\sigma_+)^{1-\beta}-(\sigma_-)^{1-\beta}}{ 2m_1^{\alpha}m_2^{\beta}(1-\beta)\sqrt{\sigma^2-1}}\,,\\[5pt]
\mathcal{I}^{(1)}(\alpha,\beta)
&=0\,,
\\
\begin{split} 
\mathcal{I}^{(2)}(\alpha,\beta)
&=\frac{b^2}{4m_1^{\alpha}m_2^{\beta}(\sigma^2-1)^{\frac{3}{2}}}\bigg[-(\sigma_+)^{1-\beta}\bigg(\frac{1}{1-\beta}+\frac{2\sigma\sigma_+}{\beta-2}-\frac{\sigma_+^2}{\beta-3}\bigg)
\\
&+2(\sigma_-)^{1-\beta}\bigg(\frac{1}{3-4\beta+\beta^2}-\frac{\sigma^2}{6-5\beta+\beta^2}+\frac{\sigma\sqrt{\sigma^2-1}}{6-5\beta+\beta^2}\bigg)\bigg]\,.
\end{split}
\end{align}
Furthermore it is easy to show that for the special values $\beta\to1,2,3$ the result of the integrals can be written in terms of $\mathrm{arcccosh}\sigma$:
\begin{align}
&\mathcal{I}^{(0)}(\alpha,1)=\frac{\mathrm{arccosh}\sigma}{m_1^{\alpha}m_2\sqrt{\sigma^2-1}}\,,
\\&\mathcal{I}^{(2)}(\alpha,1)=\frac{b^2}{2m_1^{\alpha}m_2(\sigma^2-1)^{\frac{3}{2}}}(\sigma\sqrt{\sigma^2-1}-\mathrm{arccosh}\sigma)\,,
\\&\mathcal{I}^{(2)}(\alpha,2)=-\frac{b^2}{m_1^{\alpha}m_2^2(\sigma^2-1)^{\frac{3}{2}}}(\sqrt{\sigma^2-1}-\sigma\mathrm{arccosh}\sigma)\,,
\\&\mathcal{I}^{(2)}(\alpha,3)=\frac{b^2}{2m_1^{\alpha}m_2^3(\sigma^2-1)^{\frac{3}{2}}}(\sigma\sqrt{\sigma^2-1}-\mathrm{arccosh}\sigma)\,.
\end{align}
\subsubsection*{Center-of-mass frame}
In this frame, the incoming momenta are parametrized as
\begin{align}
\label{CMparam}
p_{1}^{\mu}=-(E_1,0,0,p)\,,\qquad p_{2}^{\mu}=-(E_2,0,0,-p)\,,
\end{align}
with $\sqrt{s}=E_1+E_2=\sqrt{m_1^2+m_2^2+2m_1m_2\sigma}$ and
\begin{align}
E_1=\frac{m_1(m_1+m_2\sigma)}{\sqrt{s}}\,,\qquad E_2=\frac{m_2(m_2+m_1\sigma)}{\sqrt{s}}\,,\qquad p=\frac{m_1m_2\sqrt{\sigma^2-1}}{\sqrt{s}}\,.
\end{align}
Furthermore, without loss of generality, we choose the impact parameter aligned along the $x^2$ direction, $b^{\mu}=(0,b,0,0)$. Substituting in $\mathcal{I}^{(i)}(\alpha,\beta)$ and performing the integration over $\phi$ we obtain
\begin{align}
\mathcal{I}^{(j)}(\alpha,\beta)=\frac{(b)^j}{4\pi}\frac{[1+(-1)^j]^2\sqrt{\pi}\,\Gamma(\frac{1+j}{2})}{j\Gamma(\frac{j}{2})}\int_{-1}^1\frac{dx(1-x^2)^{\frac{j}{2}}}{(E_1-px)^{\alpha}(E_2+px)^{\beta}}\,.
\end{align}
Notice that, just as for the frame attached to particle 1, when $j=1$, the above integral vanishes. For $j=0,2$, it is possible to exactly perform the integration over $x$. However, for practical reasons, we do not display explicitly the results here.

\subsubsection*{Integrals for the $\omega^2\log\omega$ spectrum}
Together with the integrals discussed above, in order to compute the $\omega^2\log\omega$ contribution to the spectrum, we also need the compute family of integrals
\begin{align}
    \mathcal{J}_{1,2}(\alpha,\beta)=\int\frac{d\Omega}{4\pi}\frac{\log(-v_{1,2}\cdot n)}{(p_1\cdot n)^{\alpha}(p_2\cdot n)^{\beta}}\,,
\end{align}
that for simplicity we discuss only in the frame initially attached to particle 1. In this case using the parameterization in  \eqref{par1} we have $\mathcal{J}_{1}(\alpha,\beta)=\mathcal{I}^{(0)}(\alpha,\beta)$ and 
\begin{equation}
\begin{aligned}
\mathcal{J}_2(\alpha,\beta)
=
-\frac{1}{m_2^\beta}\frac{\partial}{\partial\beta}
\left[ 
m_2^\beta
\mathcal{I}^{(0)}(\alpha,\beta)
\right]
=\frac{\sigma_-^{1-\beta}[1+(\beta-1)\log\sigma_-]-\sigma_+^{1-\beta}[1+(\beta-1)\log\sigma_+]}{2m_1^{\alpha }m_2^{\beta}(\beta-1)^2\sqrt{\sigma^2-1}}\,.
\end{aligned}
\end{equation}

\section{From time domain to frequency domain}
\label{app:FTtomega}

In this appendix, we review how the large-$|t|$ expansion and the nonanalytic terms in the small-$\omega$ expansion are connected via Fourier transform from time to frequency domain.
The key relation to this effect is that, if a given real function $f(t)$ satisfies
\begin{equation}\label{eq:Apmdeft}
    f(t) \sim 
    A_0^{(\pm)} 
    +\sum_{n=1}^\infty \frac{A_{n}^{(\pm)}}{t^n}\,(\log |t|)^{n-1}
    + \cdots
    \,,
    \qquad
    \text{as }t\to\pm\infty
\end{equation}
then its Fourier transform
\begin{equation}\label{eq:FTdef}
    \tilde{f}(\omega)
    =
    \int_{-\infty}^{+\infty} e^{i\omega t} f(t)\,dt
\end{equation}
behaves as follows
\begin{equation}\label{eq:Apmlogs}
    \tilde{f}(\omega)
    \sim
    \frac{i}{\omega}\left(A_0^{(+)}-A_0^{(-)}\right)
    - \sum_{n=1}^\infty \left(A_n^{(+)}-A_n^{(-)}\right)(-i\omega)^{n-1}\frac{(\log\omega)^n}{n!}
     + \cdots\,,\qquad
    \text{as }\omega\to 0^+.
\end{equation}
More explicitly, if
\begin{equation}
    f(t)
    \sim
    \begin{cases}
    A + \dfrac{B}{t} + \dfrac{F}{t^2}\,\log t + \dfrac{H}{t^2}\,(\log t)^2 + \cdots
    \quad
    &\text{as }t\to +\infty
    \\
    \\
    0 + \dfrac{C}{t} + \dfrac{G}{t^3}\,\log |t| + \dfrac{I}{t^3}\,(\log |t|)^2 + \cdots
    \quad
    &\text{as }t\to-\infty
    \end{cases}
\end{equation}
then,
\begin{equation}
    \tilde{f}(\omega) 
    \sim
    \frac{iA}{\omega}
    -(B-C)\log\omega
    +\frac{i}{2}(F-G) \, \omega(\log\omega)^2
    +\frac{1}{6}(H-I) \, \omega^2(\log\omega)^3
    +\cdots
\end{equation}
and, in particular, we recover the first three terms discussed in \cite{Laddha:2018vbn,Sahoo:2018lxl,Saha:2019tub,Sahoo:2021ctw}.

To obtain \eqref{eq:Apmlogs}, we first rewrite \eqref{eq:FTdef} as
\begin{equation}
    \tilde{f}(\omega)
    =
    \tilde{f}_+(\omega)
    +
    \tilde{f}_-(\omega)\,,
    \qquad
    \tilde{f}_\pm(\omega)
    =
    \lim_{\epsilon\to0}
    \int_0^\infty
    e^{\pm i\omega t}
    t^\epsilon
    f(\pm t)\,dt\,.
\end{equation}
In the limit of small positive $\omega$, we need to expand  each term taking into account the two regions $|t|^{-1}\gg \omega\to0^+$ and $|t|^{-1}\sim \omega\to0^+$. In the first region, we can expand the exponential $e^{\pm i\omega t}$, while in the second we can expand $f(t)$ by means of \eqref{eq:Apmdeft},
\begin{equation}
    \begin{split}
        \int_0^\infty
    e^{i\omega t}
    t^\epsilon
    f(t)\,dt
    &=
    \sum_{n=0}^\infty
    \frac{(i\omega)^n}{n!}
    \int_0^\infty t^{n+\epsilon} f(t)\, dt
    \\
    &+
    A_0^{(+)}
    \int_0^\infty t^\epsilon e^{i\omega t} dt
    +
    \sum_{n=0}^\infty 
    A_0^{(+)}
    \frac{d^{n}}{d\epsilon^n}
    \int_0^\infty 
    t^{-n+\epsilon} 
    e^{i\omega t}\,\frac{dt}{t}
    +\cdots
    \end{split}
\end{equation}
where the $\cdots$ stand for subleading logarithms. Evaluating the integrals in the last line by giving $\omega$ a small positive imaginary part, $\omega_+\equiv \omega+i0$,
\begin{equation}
    \begin{split}
    \int_0^\infty
    e^{i\omega t}
    t^\epsilon
    f(t)\,dt
    &=
    \sum_{n=0}^\infty
    \frac{(i\omega)^n}{n!}
    \int_0^\infty t^{n+\epsilon} f(t)\, dt
    \\
    &+
    A_0^{(+)}
    \frac{\Gamma(1+\epsilon)}{-i\omega_+}
    +
    \sum_{n=0}^\infty 
    A_0^{(+)}
    \frac{d^{n}}{d\epsilon^n}
    \frac{\Gamma(\epsilon-n)}{(-i\omega_+)^{\epsilon-n}}
    +\cdots
    \end{split}
\end{equation}
In the limit $\epsilon\to0$, the first line will develop singularities canceling the poles arising from the sum in the second line. However, the leading logarithms are manifestly finite and come from the simple pole of $\Gamma(\epsilon-n)$,
\begin{equation}
    (-i\omega_+)^n \frac{d^n}{d\epsilon^n} \frac{(-1)^n}{n!\epsilon}\frac{(-\epsilon)^{n+1}}{(n+1)!}(\log\omega_+)^{n+1} = - (-i\omega_+)^n \frac{(\log\omega_+)^{n+1}}{(n+1)!}
\end{equation}
and thus
\begin{equation}\label{eq:fplus}
    \tilde{f}_+(\omega)
    \sim
    \frac{i}{\omega_+}\,A_0^{(+)}
    - \sum_{n=1}^\infty 
    (-i\omega_+)^{n-1}\frac{(\log\omega_+)^n}{n!}
    \,A_n^{(+)}
     + \cdots\,,\quad
    \text{as }\omega\to 0^+.
\end{equation}
Sending $A_n^{(+)}\to (-1)^n A_n^{(-)}$ to take into account that $(-t)^n=(-1)^n t^n$, and $\omega_+\to -\omega_-$ with $\omega_-=\omega-i0$,
we similarly obtain
\begin{equation}\label{eq:fminus}
    \tilde{f}_-(\omega)
    \sim
    -\frac{i}{\omega_-}\,A_0^{(-)}
    + \sum_{n=1}^\infty 
    (-i\omega_-)^{n-1}\frac{(\log\omega_-)^n}{n!}
    \,A_n^{(-)}
     + \cdots\,,\quad
    \text{as }\omega\to 0^+.
\end{equation}
Combining the last two relations, and neglecting the $\pm i0$ prescriptions, we recover \eqref{eq:Apmlogs}.

When taking the inverse Fourier transform, it is convenient to deform the integration contour as indicated by the $\pm i0$ in order to recover \eqref{eq:Apmdeft} from \eqref{eq:fplus}, \eqref{eq:fminus} (see Appendix C of \cite{Sahoo:2020ryf}).
In general, the basic Fourier transform is as follows, for $n\ge k\ge 1$,
\begin{equation}
I_{n,k}(t)
=
-\int_{-\infty}^{+\infty}
e^{-i\omega t} \frac{(-i\omega)^{n-1}}{n!}\,(\log\omega_+)^{n-k} (\log\omega_-)^k\,\frac{d\omega}{2\pi}
\end{equation}
and one finds \cite{Sahoo:2020ryf}
\begin{equation}\label{eq:basicPM}
I_{n,k}(t) 
\sim 
\begin{cases}
\dfrac{n-k}{n}\,\dfrac{(\log t)^{n-1}}{t^n} & \text{as }t\to+\infty
\\
&
\\
-\dfrac{k}{n}\,\dfrac{(\log |t|)^{n-1}}{t^n} & \text{as }t\to-\infty
\end{cases}
\end{equation}
to leading log accuracy.
Using these relations, together with
\begin{equation}
     \int_{-\infty}^{+\infty} e^{-i\omega t}\frac{i}{\omega_+}\,\frac{d\omega}{2\pi} = 
     \int_{-\infty}^{+\infty} \frac{e^{i\omega t}}{\omega-i 0}\,\frac{d\omega}{2i\pi} = \theta(t)\,,
\end{equation}
and expanding the exponential series, one finds that if
\begin{equation}\label{eq:resum2omegaPM}
\tilde{f}(\omega) \sim i\left(\frac{c_+}{\omega_+}-\frac{c_-}{\omega_-}\right)\,\omega_+^{-ia \omega}\,\omega_-^{-ib\omega} + \cdots
\end{equation}
then
\begin{equation}\label{eq:resum2tPM}
\begin{split}
f(t) &\sim  \left(
c_+ - \frac{(c_+-c_-)a}{t-(a+b)\log t}
\right)+\cdots \qquad \text{as }t\to+\infty\,,
\\
f(t) &\sim \left(
c_- +\frac{(c_+-c_-)b}{t-(a+b)\log |t|} \right)+\cdots\qquad \text{as }t\to-\infty
\end{split}
\end{equation}
where both expressions capture the leading logarithms in the small-$\omega$/large-$t$ expansions.

\section{Detailed derivation of the equations in~\ref{PMwaveform} }
\label{AppC}

In this appendix, we give some details on how to derive the results presented in Subsection~\ref{PMwaveform}. We focus on the $\omega^{-1}$, $\log \omega$ and $\omega (\log \omega)^2$ contributions to the waveform. Starting from \eqref{resummedWeinb} and writing explicitly the various terms we get (see also \cite{Krishna:2023fxg,Alessio:2024wmz})
\begin{align}
 \tilde{w}^{\mu\nu}
 &=-i\bigg[\hat{\mathcal{S}}^{\mu\nu}_{(0)}+\bigg(-\mathcal{W}_{\mathrm{phase}}\hat{\mathcal{S}}^{\mu\nu}_{(0)}+\hat{\mathcal{S}}_{(1)}^{\mu\nu}\mathcal{W}_{\mathrm{reg}}\bigg)+\bigg(\frac{1}{2}\mathcal{W}^2_{\mathrm{phase}}\hat{\mathcal{S}}^{\mu\nu}_{(0)}-\mathcal{W}_{\mathrm{phase}}\hat{\mathcal{S}}^{\mu\nu}_{(1)}(\mathcal{W}_{\mathrm{reg}})
 \\
 &+\frac{1}{2}\sum_{a=1}^4\frac{k_{\rho}k_{\sigma}}{p_a\cdot k}(\hat{J}_a^{\mu\rho}\mathcal{W}_{\mathrm{reg}})(\hat{J}_a^{\nu\sigma}\mathcal{W}_{\mathrm{reg}})\bigg)\bigg]=\tilde{w}^{\mu\nu}_{[\omega^{-1}]}\omega^{-1}+\tilde{w}^{\mu\nu}_{[\log\omega]}\log\omega+\tilde{w}^{\mu\nu}_{[\omega(\log\omega)^2]}\omega(\log\omega)^2\,,
 \nonumber
\end{align}
where the soft operators are
\begin{equation}
\hat{\mathcal{S}}^{\mu\nu}_{(0)} =\sum_{a=1}^4 \frac{p_a^\mu p_a^\nu}{p_a\cdot k},\qquad   \hat{\mathcal{S}}_{(1)}^{\mu\nu}= \frac{1}{2}\sum_{a=1}^4 \frac{k_\rho}{p_a\cdot k} (p_a^\mu \hat{J}_a^{\nu \rho} + p_a^\nu \hat{J}_a^{\mu \rho}),
    \label{C2}
\end{equation}
and 
\begin{align}
\label{C4}
&\mathcal{W}_\text{reg} = -i G m_1 m_2 \log \omega \Bigg( \frac{2\sigma_{12}^2 - 1}{\sqrt{\sigma_{12}^2 -1}}+  \frac{2\sigma_{34}^2 - 1}{\sqrt{\sigma_{34}^2 -1}} \Bigg),\\&\mathcal{W}_\text{phase}=-2iG (p_1+p_2)\cdot k \log \omega.
\end{align}
Then one gets, using $k^\mu = \omega n^\mu$ 
\begin{align}
&  \tilde{w}^{\mu \nu}_{[\omega^{-1}]}= iA_0^{\mu\nu}(p_i,p_f) \, , 
\label{C5}\\
\nonumber& \tilde{w}^{\mu \nu}_{[\log \omega]}= - 2G[(p_1+p_2)\cdot n ]A_0^{\mu\nu}(p_i,p_f)+
G\bigg[\frac{\sigma_{12}(2\sigma_{12}^2- 3)}{(\sigma^2_{12}-1)^{3/2}}B^{\mu \nu}(p_i)  
\\&+\frac{\sigma_{34}(2\sigma_{34}^2- 3)}{(\sigma^2_{34}-1)^{3/2}}B^{\mu \nu}(p_f)\bigg], \label{C6} \\
& \nonumber\tilde{w}^{\mu \nu}_{[\omega (\log \omega)^2]}=\frac{i}{2}\bigg\{-[2G(p_1+p_2)\cdot n]^2A_0^{\mu\nu}(p_i,p_f)+4G^2[(p_1+p_2)\cdot n]\bigg[\frac{\sigma_{12}(2\sigma_{12}^2- 3)}{(\sigma^2_{12}-1)^{3/2}}B^{\mu \nu}(p_i)  \\\nonumber &+\frac{\sigma_{34}(2\sigma_{34}^2- 3)}{(\sigma^2_{34}-1)^{3/2}}B^{\mu \nu}(p_f)\bigg]+G^2\bigg[[(p_1+p_2)\cdot n]\bigg(\frac{\sigma_{12}(2\sigma_{12}^2- 3)}{(\sigma^2_{12}-1)^{3/2}}\bigg)^2B^{\mu \nu}(p_i)\\&+[(p_3+p_4)\cdot n]\bigg(\frac{\sigma_{34}(2\sigma_{34}^2- 3)}{(\sigma^2_{34}-1)^{3/2}}\bigg)^2B^{\mu \nu}(p_f)\bigg]\bigg\},
    \label{C7}
\end{align}
where $A_0^{\mu\nu}$ and $B^{\mu\nu}$ are given in \eqref{eq:A0} and \eqref{Bmunu} and $p_i=(p_1,p_2)$, $p_f=(p_3,p_4)$. As already explained in section \ref{PMwaveform}, the PM expansion of the soft waveform can be achieved by re-expressing it in terms of ${\tilde{p}}_1^{\mu}, {\tilde{p}}_2^{\mu}$ introduced in \eqref{A} and then using the PM expansion of the two classical impulses $Q_1^{\mu}, Q_2^{\mu}$ given in \eqref{PMmomenta1} and \eqref{PMmomenta}.
Here we want to perform the calculation at two-loop by selecting the terms that are proportional to $G^3$. For the 3PM $\log\omega$ and $\omega(\log\omega)^2$ waveforms we can safely take $Q_1^{\mu} =- Q_2^{\mu} =Q^{\mu}$
and  $\sigma_{12}= \sigma_{34} \equiv \sigma$ because, at such order, we only need $Q^{\mu}_{\mathrm{1PM}}$ and $Q^{\mu}_{\mathrm{2PM}}$. We introduce the notation
\begin{align}
&B^{\mu\nu}(p_i)=\sum_{n=0}^{\infty}{}_iB^{\mu\nu}_{\alpha_1\cdots\alpha_n}(\tilde{p}_i)Q^{\alpha_1}\cdots Q^{\alpha_n}\equiv \sum_{n=0}^{\infty}{}_iB^{\mu\nu}_{n}(\tilde{p}_i),\\&B^{\mu\nu}(p_f)=\sum_{n=0}^{\infty}{}_fB^{\mu\nu}_{\alpha_1\cdots\alpha_n}(\tilde{p}_i)Q^{\alpha_1}\cdots Q^{\alpha_n}\equiv \sum_{n=0}^{\infty}{}_fB^{\mu\nu}_{n}(\tilde{p}_i).
\end{align}
We find
\begin{align}
&  {}_iB_0^{\mu \nu}(\tilde{p}_i)= {}_fB_0^{\mu \nu} (\tilde{p}_i)=  {\tilde{p}}_1^\mu {\tilde{p}}_1^\nu \frac{{\tilde{p}}_2\cdot n}{{\tilde{p}}_1\cdot k}   +{\tilde{p}}_2^\mu {\tilde{p}}_2^\nu \frac{{\tilde{p}}_1\cdot n}{{\tilde{p}}_2\cdot n}  -{\tilde{p}}_1^\mu {\tilde{p}}_2^\nu-{\tilde{p}}_1^\nu {\tilde{p}}_2^\mu \, , \label{C16} \\
& {}_iB_1^{\mu \nu}(\tilde{p}_i)= - {}_fB_1^{\mu \nu} (\tilde{p}_f)= \frac{ ({\tilde{p}}_1+ {\tilde{p}}_2)\cdot n}{2} \Bigg[ \bigg(\frac{{\tilde{p}}_1^\mu{\tilde{p}}_1^\nu}{({\tilde{p}}_1 \cdot n)^2} - \frac{{\tilde{p}}_2^\mu{\tilde{p}}_2^\nu}{({\tilde{p}}_2 \cdot n)^2} \bigg) (Q\cdot n) - \frac{{\tilde{p}}_1^\mu Q^\nu + {\tilde{p}}_1^\nu Q^\mu}{{\tilde{p}}_1\cdot n} \nonumber \\
& + \frac{{\tilde{p}}_2^\mu Q^\nu + {\tilde{p}}_2^\nu Q^\mu}{{\tilde{p}}_2\cdot k}\Bigg] \, ,\label{C17}\\
& {}_iB_2^{\mu \nu}(\tilde{p}_i)=  {}_f B_2^{\mu \nu} (\tilde{p}_f)= \frac{({\tilde{p}}_1+ {\tilde{p}}_2)\cdot n}{4}\Bigg[ \bigg( \frac{{\tilde{p}}_1^\mu {\tilde{p}}_1^\nu}{({\tilde{p}}_1\cdot n)^3} + \frac{{\tilde{p}}_2^\mu {\tilde{p}}_2^\nu}{({\tilde{p}}_2\cdot n)^3}\bigg) (Q\cdot n) -
\frac{{\tilde{p}}_1^\mu Q^\nu + {\tilde{p}}_1^\nu Q^\mu }{({\tilde{p}}_1\cdot n)^2} \nonumber \\
& -
\frac{{\tilde{p}}_2^\mu Q^\nu + {\tilde{p}}_2^\nu Q^\mu  }{({\tilde{p}}_2\cdot n)^2}\Bigg]( Q\cdot n) + \frac{[({\tilde{p}}_1+ {\tilde{p}}_1)\cdot n]^2}{4 ({\tilde{p}}_2\cdot n)({\tilde{p}}_2\cdot  n)} Q^\mu Q^\nu \,.
    \label{C18}
\end{align}
Similarly, for the $\omega^{-1}$ contribution to the waveform we expand $A_{-1}^{\mu\nu}$ as
\begin{align}
A_{-1}^{\mu\nu}(p_i,p_f)=\sum_{n=1}^{\infty}A_{-1\,n}^{\mu\nu}(\tilde{p}_i).
\end{align}
However, for the 3PM $\omega^{-1}$ waveform, and in particular for the term $A_{-1\,1}^{\mu\nu}(\tilde{p}_i)$, we need to distinguish between $Q_1^{\mu}$ and $Q_2^{\mu}$. This is related to the presence of the non-linear memory as discussed in \ref{PMwaveform}. We find
\begin{align}
&\nonumber A_{-1\,1}^{\mu \nu}(\tilde{p}_i)=\frac{1}{{\tilde{p}}_1\cdot n}\Bigg[ \bigg({\tilde{p}}_1^\mu Q_1^\nu + {\tilde{p}}_1^\nu Q_1^\mu \bigg) -{\tilde{p}}_1^\mu {\tilde{p}}_1^\nu  \frac{Q_1\cdot n}{{\tilde{p}}_1\cdot n} \Bigg]+\frac{1}{{\tilde{p}}_2 \cdot n}\Bigg[ \bigg({\tilde{p}}_2^\mu Q_2^\nu + {\tilde{p}}_2^\nu Q_2^\mu \bigg) \\&-{\tilde{p}}_2^\mu {\tilde{p}}_2^\nu  \frac{Q_2\cdot n}{{\tilde{p}}_2\cdot n} \Bigg]\,,\label{C19} \\
& A_{-1\,2}^{\mu\nu}(\tilde{p}_i)=0,\\
& A_{-1\,3}^{\mu \nu}(\tilde{p}_i)= \frac{1}{{\tilde{p}}_1\cdot n} \Bigg[ \bigg({\tilde{p}}_1^\mu Q^\nu + {\tilde{p}}_1^\nu Q^\mu  -{\tilde{p}}_1^\mu {\tilde{p}}_1^\nu  \frac{Q\cdot n}{{\tilde{p}}_1\cdot n} \bigg)\left( \frac{Q\cdot n}{ {2\tilde{p}}_1\cdot n}\right)^{2} - \frac{Q^\mu Q^\nu}{4} \frac{Q\cdot n}{ {\tilde{p}}_1\cdot n} \Bigg]\nonumber \\
 & - \frac{1}{{\tilde{p}}_2\cdot n}  \Bigg[\bigg({\tilde{p}}_2^\mu Q^\nu + {\tilde{p}}_2^\nu Q^\mu  -{\tilde{p}}_2^\mu {\tilde{p}}_2^\nu  \frac{Q\cdot n}{{\tilde{p}}_2\cdot n} \bigg)\left( \frac{Q\cdot n}{ {2\tilde{p}}_2\cdot n}\right)^{2} - \frac{Q^\mu Q^\nu}{4} \frac{Q\cdot n}{ {\tilde{p}}_2\cdot n} \Bigg] \, . 
    \label{C20}
\end{align}
As clear from \eqref{C6}, the 3PM $\log \omega$ term receives a contribution only from $B_0^{\mu \nu}(\tilde{p}_i)$, $B_{2}^{\mu \nu}(\tilde{p}_i)$ and from $A_{-1\,1}^{\mu \nu}(\tilde{p}_i)$, that we rewrite here in terms of the variables introduced in \eqref{A} and \eqref{C}:
\begin{align}
& {}_iB_0^{\mu \nu}(\tilde{p}_i)={}_fB_0^{\mu \nu}(\tilde{p}_i) = \frac{{\tilde{m}}_1 {\tilde{m}}_2}{{\tilde{\alpha}}_1 {\tilde{\alpha}}_2} \Bigg[{\tilde{u}}_1^\mu {\tilde{u}}_1^\nu ({\tilde{\alpha}}_2)^2 +{\tilde{u}}_2^\mu {\tilde{u}}_2^\nu ({\tilde{\alpha}}_1)^2 -{\tilde{\alpha}}_1 {\tilde{\alpha}}_2 ({\tilde{u}}_1^\mu {\tilde{u}}_2^\nu+ {\tilde{u}}_1^\nu {\tilde{u}}_2^\mu )\Bigg]\, , \label{C21} \\
& {}_iB_2^{\mu \nu}(\tilde{p}_i)={}_fB_2^{\mu \nu}(\tilde{p}_i)=\frac{{\tilde{m}}_1 {\tilde{\alpha}}_1 +{\tilde{m}}_2 {\tilde{\alpha}}_2}{4{\tilde{m}}_1 {\tilde{m}}_2 {\tilde{\alpha}}_1^3 {\tilde{\alpha}}_2^3 }
\Bigg[ \bigg({\tilde{m}}_2 {\tilde{\alpha}}_2^3 {\tilde{u}}_1^\mu {\tilde{u}}_1^\nu +{\tilde{m}}_1 {\tilde{\alpha}}_1^3 {\tilde{u}}_2^\mu {\tilde{u}}_2^\nu
\bigg)(Q\cdot n)^2 -  {\tilde{m}}_2 {\tilde{\alpha}}_2^3 {\tilde{\alpha}}_1 ({\tilde{u}}_1^\mu Q^\nu \nonumber\\& + {\tilde{u}}_1^\nu Q^\mu)(Q\cdot n)  -  {\tilde{m}}_1 {\tilde{\alpha}}_1^3 {\tilde{\alpha}}_2 ({\tilde{u}}_2^\mu Q^\nu + {\tilde{u}}_2^\nu Q^\mu)(Q\cdot n) +({\tilde{m}}_1 {\tilde{\alpha}}_1 +{\tilde{m}}_2 {\tilde{\alpha}}_2)
{\tilde{\alpha}}_1^2 {\tilde{\alpha}}_2^2 Q^\mu Q^\nu\Bigg] \, , \label{C22}\\
\nonumber & A_{-1\,1}^{\mu \nu}(\tilde{p}_i)=
 -\frac{1}{{\tilde{\alpha}}_1^2 {\tilde{\alpha}}_2^2} \Bigg[ {\tilde{\alpha}}_2^2(Q_1\cdot n) {\tilde{u}}_1^{\mu}\tilde{u}_1^{\nu}+{\tilde{\alpha}}_1^2(Q_2\cdot n) \tilde{u}_2^{\mu}\tilde{u}_2^{\nu}+ 
{\tilde{\alpha}}_2^2 {\tilde{\alpha}}_1( \tilde{u}_1^{\mu}Q_1^{\nu}+\tilde{u}_1^{\nu}Q_1^{\mu})\\& +
{\tilde{\alpha}}_2 {\tilde{\alpha}}_1^2( \tilde{u}_2^{\mu}Q_2^{\nu}+\tilde{u}_2^{\nu}Q_2^{\mu}) \Bigg]\label{C19bis}.
\end{align}
By saturating the previous quantities with the  polarisation $\varepsilon^{\mu\nu}=\varepsilon^{\mu}\varepsilon^{\nu}$ satisfying $\varepsilon^2=0=\varepsilon\cdot n$ and denoting by $T_n\equiv \varepsilon_{\mu}T_n^{\mu\nu}\varepsilon_{\nu}$ for a generic tensor $T_n^{\mu\nu}$ we obtain, 
\begin{align}
&  {}_iB_0(\tilde{p}_i)= \frac{{\tilde{m}}_1 {\tilde{m}}_2}{{\tilde{\alpha}}_1 {\tilde{\alpha}}_2} g_1 
,\label{C25} \\
& {}_iB_2(\tilde{p}_i)=
\frac{{\tilde{m}}_1 {\tilde{\alpha}}_1 +{\tilde{m}}_2 {\tilde{\alpha}}_2}{4{\tilde{m}}_1 {\tilde{m}}_2 {\tilde{\alpha}}_1^3 {\tilde{\alpha}}_2^3 } \frac{Q_\text{1PM}^2}{b_e^2} g_5 \, ,
\label{C26} \\
&A_{-1\,1}(\tilde{p}_i)=- \frac{Q_\text{2PM}}{b_e({\tilde{\alpha}}_1 {\tilde{\alpha}}_2)^2}g_4 \, ,
    \label{C27}
\end{align} 
and hence for the 3PM $\log\omega$ waveform 
\begin{align}
& \tilde{w}^{[\log \omega]}_{3\mathrm{PM}} = 
 2 G 
\Bigg\{ ({\tilde{m}}_1{\tilde{\alpha}}_1+{\tilde{m}}_2{\tilde{\alpha}}_2)\Bigg[ \frac{Q_\text{1PM}^2}{4b_e^2 {\tilde{m}}_1 {\tilde{m}}_2{\tilde{\alpha}}_1^3 {\tilde{\alpha}}_2^3 } \frac{\sigma (2\sigma^2-3)}{(\sigma^2-1)^{3/2}} g_5 - \frac{Q_\text{2PM}}{ {b_e \tilde{\alpha}}_1^2 {\tilde{\alpha}}_2^2} g_4\Bigg] \nonumber \\
& +  \frac{\sigma (2\sigma^2-3)}{(\sigma^2-1)^{3/2}}\frac{(m_1^2 +m_2^2)Q_\text{1PM}^2}{8{\tilde{\alpha}}_1 {\tilde{\alpha}}_2 m_1 m_2 } g_1\Bigg\} \,\,,
    \label{C28}
\end{align}
in terms of the structures in \eqref{gstructures}.
The first line  reproduces \eqref{F} after using  the relation ${\tilde{m}}_1 {\tilde{\alpha}}_1 +{\tilde{m}}_2 {\tilde{\alpha}}_2= E$. The last term comes from the tree level $\log\omega$ waveform  by expressing ${\tilde{m}}_1, {\tilde{m}}_2$ in terms of $m_1, m_2$ obtaining an additional power of $Q^2$ using equation \eqref{masses}. Let us now compute the 3PM $\omega(\log\omega)^2$ waveform. We need to express ${}_{i,f}B_1^{\mu\nu}(\tilde{p}_i)$ in \eqref{C17} in terms of the variables in \eqref{A} and \eqref{C}, that gives
\begin{align}
& {}_iB_{1}^{\mu \nu}(\tilde{p}_i) = \frac{{\tilde{m}}_1 {\tilde{\alpha}}_1 + {\tilde{m}}_2 {\tilde{\alpha}}_2}{2b_e ({\tilde{\alpha}}_1 {\tilde{\alpha}}_1)^2} Q_\text{1PM}
\nonumber \\
& \times\Bigg[ \bigg( {\tilde{\alpha}}_2^2 {\tilde{u}}_1^\mu {\tilde{u}}_1^\nu- {\tilde{\alpha}}_1^2 {\tilde{u}}_2^\mu {\tilde{u}}_2^\nu\bigg)(b_e\cdot n) + {\tilde{\alpha}}_1 {\tilde{\alpha}}_2^2 ({\tilde{u}}_1^\mu b_e^\nu+ {\tilde{u}}_1^\nu b_e^\mu)- {\tilde{\alpha}}_1^2 {\tilde{\alpha}}_2 ({\tilde{u}}_2^\mu b_e^\nu+ {\tilde{u}}_2^\nu b_e^\mu) \Bigg].
    \label{C29}
\end{align}
Saturating with the polarization vectors we find 
\begin{align}
&{}_iB_{1}(\tilde{p}_i)= -\frac{({\tilde{m}}_1 {\tilde{\alpha}}_1 + {\tilde{m}}_2 {\tilde{\alpha}}_2)Q_\text{1PM}}{2b_e ({\tilde{\alpha}}_1 {\tilde{\alpha}}_1)^2} g_4  \, .
    \label{C30}
\end{align}
Using this equation together with \eqref{C27} (with $Q_\text{1PM}$  instead of $Q_\text{2PM}$) in \eqref{C7}, we get  
\begin{align}
& \tilde{w}^{[\omega (\log \omega)^2]}_{3\mathrm{PM}} = -2i G^2 (m_1{\tilde{\alpha}}_1+m_2{\tilde{\alpha}}_2)^2\bigg[ 1 + \frac{1}{4}\left(\frac{\sigma (2\sigma^2-3)}{(\sigma^2-1)^{3/2}} \right)^2\bigg]\frac{Q_\text{1PM}}{b_e {\tilde{\alpha}}_1^2 {\tilde{\alpha}}_2^2} g_4 .
    \label{C31}
\end{align}
Let us now conclude with the $\omega^{-1}$ term. We saturate \eqref{C20} with the polarisation vectors and we obtain
\begin{align}
& A_{-1\,3}(\tilde{p}_i)= -\frac{(b_e\cdot n)}{4 {\tilde{m}}_1^2 {\tilde{m}}_2^2 {\tilde{\alpha}}_1^4 {\tilde{\alpha}}_2^4} \left( \frac{Q_\text{1PM}}{b_e}\right)^3 \Bigg[ (b_e\cdot  n)^2 \bigg({\tilde{m}}_1^2{\tilde{\alpha}}_1^4 ({\tilde{u}}_2 \cdot \varepsilon)^2- {\tilde{m}}_2^2{\tilde{\alpha}}_2^4 ({\tilde{u}}_1\cdot  \varepsilon )^2\bigg)\nonumber \\
& +2 (b_e\cdot n){\tilde{\alpha}}_1 {\tilde{\alpha}}_2 (b_e \cdot \varepsilon)\bigg(
{\tilde{m}}_1^2 {\tilde{\alpha}}_1^3 ({\tilde{u}}_2\cdot \varepsilon)-{\tilde{m}}_2^2 {\tilde{\alpha}}_2^3 ({\tilde{u}}_1\cdot  \varepsilon) \bigg)+ (b_e\cdot  \varepsilon)^2 {\tilde{\alpha}}_1^2 {\tilde{\alpha}}_2^2 \bigg({\tilde{\alpha}}_1^2{\tilde{m}}_1^2 -
{\tilde{\alpha}}_2^2{\tilde{m}}_2^2 \bigg)\Bigg] \nonumber \\
&= -\frac{(b_e\cdot  n)}{4 {\tilde{m}}_1^2 {\tilde{m}}_2^2 {\tilde{\alpha}}_1^4 {\tilde{\alpha}}_2^4} \left( \frac{Q_\text{1PM}}{b_e}\right)^3 g_3 \, .
    \label{C32}
\end{align}
Another term is obtained from \eqref{C19bis} with the 3PM momenta, including their radiative contributions in \eqref{radmom}. In total, we get for the 3PM $\omega^{-1}$ waveform

\begin{align}
& \tilde{w}^{[ \omega^{-1}]}_{3\mathrm{PM}}=\frac{iQ_\text{3PM}}{b_e({\tilde{\alpha}}_1 {\tilde{\alpha}}_2)^2} g_4 +\frac{i(b_e n)}{4 {\tilde{m}}_1^2 {\tilde{m}}_2^2 {\tilde{\alpha}}_1^4 {\tilde{\alpha}}_2^4} \left( \frac{Q_\text{1PM}}{b_e}\right)^3 g_3 - \frac{i G^3 {\tilde{m}}_1^2 {\tilde{m}}_2^2 {\cal{E}}(\sigma)}{b_e^3(\sigma^2-1) {\tilde{\alpha}}_1^2 {\tilde{\alpha}}_2^2} g_2.
\label{eq:A2loopPDV}
\end{align}
that, after the addition of the non-linear memory term, reproduces \eqref{L}.

In order to check the leading PN limit of the two-loop expressions 
for $\frac{1}{\omega}$ and $\log \omega$, given in this appendix,  with those obtained in Section~\ref{NewtonianLimit} we need to go from the tilded to the non-tilded variables as discussed in \eqref{masses} and \eqref{tildesigma}. In terms of the tilded variables we get
\begin{equation}
({\tilde{u}}_1-{\tilde{u}}_2)\cdot\varepsilon = \frac{{\tilde{p}} ({\tilde{m}}_1+{\tilde{m}}_2 )}{{\tilde{m}}_1 {\tilde{m}}_2} ({\hat{u}}\cdot \varepsilon)
= \sqrt{{\tilde{\sigma}}^2-1}({\hat{u}}\cdot \varepsilon)
 \frac{{\tilde{m}}_1+{\tilde{m}}_2}{\sqrt{{\tilde{m}}_1^2 + {\tilde{m}}_2^2 +2 {\tilde{m}}_1 {\tilde{m}}_2 {\tilde{\sigma}}}}
    \label{C39}
\end{equation}
where ${\hat{u}}^2=-1$ and we have used the expression of 
$\tilde{p}$ in the tilded c.o.m frame in terms of ${\tilde{\sigma}}$.
The ratio in the last term can be put to $1$ at the leading PN order  and then, for $\sigma \sim 1$, one gets 
 \begin{equation}
({\tilde{u}}_1-{\tilde{u}}_2)\cdot\varepsilon = {\hat{u}}\cdot\varepsilon \bigg( p_\infty - \frac{Q^2(m_1+m_2)^2}{8 p_\infty m_1^2 m_2^2} \bigg) \sim
{\hat{u}}\cdot\varepsilon \bigg( p_\infty - \frac{G^2 (m_1+m_2)^2}{2 b_e^2 p_\infty^3}\bigg) \, ,
    \label{C40}
\end{equation}
where $p_\infty= \sqrt{\sigma^2-1}$ and in the last step we have inserted $Q^\mu =- \frac{b_e^\mu}{b_e} Q_\text{1PM}$. When we insert this in the tree level terms in \eqref{eq:treelevelwsummary}, we generate  terms of order $G^3$ that must be included to agree with the Newtonian approach of Section~\ref{NewtonianLimit}. 

\section{Newtonian trajectory to leading log accuracy}
\label{app:newt}

In this appendix, we provide explicit expressions for the Newtonian-level trajectory which serve as intermediate steps to obtain the results of Subsection~\ref{NewtonianLimit}. We start by substituting the approximate solution $x \simeq |y| + 2 e_r^{-1}\log x$ in \eqref{eq:wysolvev} into the Keplerian trajectory \eqref{eq:Kepleriantrajectory}, which leads to
\begin{subequations}
\begin{align}
    x_1 &\simeq \bar{a}_r \left[ 
    e_r - \frac{|y|}{2} -\frac{1}{e_r} \log\left(|y| + \frac{2}{e_r}\log|y|\right)
    \right],\\
    x_2 &\simeq \frac{\bar{a}_r}{e_r}\,\sqrt{e_r^2-1} \left[ 
    \frac{e_r}{2}\,y \pm \log\left(|y| + \frac{2}{e_r}\log|y|\right)
    \right].
\end{align}
\end{subequations}
The next step in order to substitute into the quadrupole formula \eqref{eq:quadrupole} is to take derivatives with respect to $x^0$ and evaluate the trajectory as a function of retarded time $x^0\mapsto t$. 
In this way, retaining only the leading logarithmic dependence for large $|y|$, we find the following expressions in terms of the variable $\tau = \bar{a}_r^{-3/2} |t|+\log|t|$ introduced in \eqref{eq:tau},
\begin{equation}
    x_1 \simeq \bar{a}_r \left(
    e_r - \frac{\tau}{e_r}
    \right),\qquad
    x_2 \simeq \pm \frac{\bar{a}_r}{e_r}\,\sqrt{e_r^2-1}\ \tau\,,
\end{equation}
as well as
\begin{equation}
    \dot x_1 \simeq \mp \frac{1}{e_r\sqrt{\bar{a}_r}} \left(
    1 + \frac{1}{\tau}
    \right),\qquad
    \dot x_2 \simeq \frac{1}{e_r\sqrt{\bar{a}_r}}\,\sqrt{e_r^2-1}\left(
    1 + \frac{1}{\tau}
    \right),
\end{equation}
and
\begin{equation}
    \ddot x_1 \simeq \frac{1}{e_r\bar{a}^2_r\tau^2}\,,
    \qquad
    \ddot x_2 \simeq \mp \frac{1}{e_r\bar{a}_r^2}\,\sqrt{e_r^2-1}\,\frac{1}{\tau^2}\,.
\end{equation}
Using these expressions, we find
\begin{subequations}
\begin{align}
(x_1 x_1)^{(2)} 
&\simeq \frac{2}{e_r^2\bar{a}_r}\left(
    1 + \frac{1}{\tau}
    \right),
\\
(x_1 x_2)^{(2)} 
&\simeq \mp \frac{2\sqrt{e_r^2-1}}{e_r^2\bar{a}_r}
\left(
    1 + \frac{1}{\tau}
    \right),
\\
(x_2 x_2)^{(2)}
&\simeq
\frac{2(e_r^2-1)}{e_r^2\bar{a}_r}\left(
    1 + \frac{1}{\tau}
    \right),
\end{align}
\end{subequations}
which, via the quadrupole formula \eqref{eq:quadrupole}, yield the Newtonian expression for the waveform given in \eqref{eq:ewededuced}.

\providecommand{\href}[2]{#2}\begingroup\raggedright\endgroup

\end{document}